\documentclass[12pt,a4j]{article}
\usepackage[dvips]{graphicx}
\usepackage{enumerate}
\usepackage{amsmath}
\usepackage{amsfonts}

\begin{document}
\baselineskip=5.5mm
\centerline{\bf The multi-component \bf modified nonlinear Schr\"odinger system }\par
\centerline{\bf  with nonzero boundary conditions}\par
\bigskip
\centerline{Yoshimasa Matsuno\footnote{{\it E-mail address}: matsuno@yamaguchi-u.ac.jp}}\par

\centerline{\it Division of Applied Mathematical Science,}\par
\centerline{\it Graduate School of Science and Technology for Innovation} \par
\centerline{\it Yamaguchi University, Ube, Yamaguchi 755-8611, Japan} \par
\bigskip
\bigskip
\leftline{\bf Abstract}\par
\noindent  In a previous paper (Matsuno 2011 {\it J. Phys. A: Math. Theore.} {\bf 44} 495202),  we have presented  a
 determinantal expression of the bright $N$-soliton solution for a multi-component
 modified nonlinear Schr\"odinger (NLS)  system  with zero boundary conditions.
The present paper provides  the dark $N$-soliton solution for the same system of equations with  plane-wave boundary conditions, 
as well as the bright-dark $N$-soliton solution with mixed zero and plane-wave boundary conditions.
The proof of the $N$-soliton solutions is performed by means of an elementary theory of determinants in which Jacobi's identity
plays the central role.
 The $N$-soliton formulas obtained here include as special cases the existing soliton solutions of the NLS and derivative NLS equations
 and their integrable multi-component analogs. The new features of soliton solutions are discussed. In particular, it is 
 shown that  the $N$ constraints must be imposed on the amplitude parameters of solitons in  constructing the dark $N$-soliton solution with
 plane-wave boundary conditions.  This makes it difficult to analyze the solutions as the number of components increases.
 For  mixed type boundary conditions, the structure of soliton solutions is found to be more explicit than that of dark soliton solutions
since no constraints are imposed on the soliton parameters. Last, we comment on an integrable multi-component system associated with the
first negative flow of the multi-component derivative NLS hierarchy.
 \par
\bigskip
\bigskip
\bigskip
\noindent {\it PACS:}\ 05.45.Yv; 42.81.Dp; 02.30.Jr \par
\noindent{\it Keywords:} Modified nonlinear Schr\"odinger system; plane-wave boundary conditions; dark $N$-soliton solution \par

\newpage
\leftline{\bf  1. Introduction} \par
\bigskip
\noindent The  nonlinear Schr\"odinger (NLS)  equation is a milestone in the theory of nonlinear waves. One remarkable
feature is that it is a completely integrable equation, allowing us to solve it  by means of several exact methods of solution such as the inverse
scattering transform (IST) [1-3], B\"acklund-Darboux (BD) transformation [4-6] and Hirota's direct method [7-9].   In recent years, much attention has been
paid to the multi-component generalizations of the NLS equation  because of
their wide applicability in many physical contexts such as nonlinear optics, nonlinear water waves and plasma physics and so on [10-12]. \par
In this paper, we consider the following multi-component system of  nonlinear partial differential equations (PDEs) which is a hybrid of the
multi-component NLS system and multi-component derivative NLS system
$${\rm i}\,q_{j,t}+q_{j,xx}+\mu\left(\sum_{s=1}^n\sigma_s|q_s|^2\right)q_j+{\rm i}\gamma\left[\left(\sum_{s=1}^n\sigma_s|q_s|^2\right)q_j \right]_x=0,\quad j=1, 2, ..., n,  \eqno(1.1)$$
where $q_j=q_j(x,t) \ (j=1, 2, ..., n)$ are  complex-valued functions of $x$ and $t$, $\mu$ and $\gamma$ are real constants, $n$ is an arbitrary positive integer,
and subscripts $x$ and $t$ appended to $q_j$ denote  partial differentiations. The coefficients $\sigma_s=\pm 1\ (s=1, 2, ..., n)$ specify the sign of the nonlinearity. 
For the multi-component NLS system   for which $\gamma=0$, we can deal with the three types of cubic nonlinearities, i.e., focusing ($\sigma_s=1,\ s=1, 2, ..., n)$,
defocusing ($\sigma_s=-1,\ s=1, 2, ..., n)$ and  mixed focusing-defocusing  $(\sigma_s=1, s=1, 2, ..., m; \sigma_s=-1, s=m+1, m+2, ..., n)$  nonlinearities with $\mu>0$, where $m$ is an
arbitrary positive integer such that $1\leq m <n$.
The special systems reduced from (1.1)  have been summarized in  a previous paper [13].
The system (1.1) has been shown to be completely integrable [14] and hence the exact methods  mentioned above can be applied to it to obtain various types of soliton  solutions.  
The properties of solutions, however,  depend  essentially on the boundary conditions.   
We emphasize that  most works have been concerned with the analysis of the system under vanishing boundary conditions at spatial infinity. 
\par
The main purpose of the present paper is to construct $N$-soliton solutions ($N$: arbitrary positive integer) of the system (1.1)  
with the following two types of  boundary conditions: \par
\noindent 1) Plane-wave boundary conditions. 
$$q_j \sim \rho_j\,{\rm exp}\,{\rm i}\left(k_jx-\omega_jt+\phi_j^{(\pm)}\right), \quad x\rightarrow \pm\infty,\quad   \quad j=1, 2, ..., n, \eqno(1.2a) $$
$$\omega_j=k_j^2-\mu\sum_{s=1}^n\sigma_s\rho_s^2+\gamma\left(\sum_{s=1}^n\sigma_s\rho_s^2\right)k_j, \quad j=1, 2, ..., n. \eqno(1.2b)$$
\noindent  2) Mixed type  boundary conditions.
  $$q_j \sim 0,\quad x\rightarrow \pm\infty,\quad j=1, 2, ..., m, \eqno(1.3a)$$
$$q_{m+j} \sim \rho_j\,{\rm exp}\,{\rm i}\left(k_jx-\omega_jt+\phi_j^{(\pm)}\right), \quad x\rightarrow \pm\infty,\quad j=1, 2, ..., n-m, \eqno(1.3b)$$
$$\omega_j=k_j^2-\mu\sum_{s=1}^{n-m}\sigma_s\rho_s^2+\gamma\left(\sum_{s=1}^{n-m}\sigma_s\rho_s^2\right)k_j, \quad j=1, 2, ..., n-m. \eqno(1.3c)$$
Here,  $\rho_j, \ k_j$ and $\phi_j^{(\pm)}$  are arbitrary real constants and $\omega_j$ from $(1.2b)$  and $(1.3c)$ represent the linear dispersion relations for components with
the plane-wave boundary conditions.  For case 1), all the components approach the specified plane waves at infinity whereas for case 2), the first $m$ components vanish at infinity and
the remaining $n-m$ components have the specified plane waves.
It is important that for the nonzero background fields like $(1.2a)$ and $(1.3b)$, each component has different wavenumber and frequency.
This causes  difficulties in constructing soliton solutions. In particular, in the case of 1) above, one must impose certain constraints between the real and
imaginary parts of the complex parameters characterizing the amplitude (or the velocity) of solitons. 
This peculiar feature has never been encountered in the case of zero boundary conditions [13].
\par
The several exact methods are now available for the nonlinear PDEs. Among them, the IST provides a general scheme 
for solving the initial value problems of a wide class of completely integrable nonlinear PDEs [1-3].  For instance, the defocusing NLS equation has been
solved by the IST [15, 16] while the analysis of the focusing  NLS equation with nonzero boundary conditions  was made quite recently in [17, 18] and used to explore the nonlinear stage of  modulational
instability.  The  IST has been developed as well for the derivative NLS equation subjected to nonzero boundary conditions [19].  The situation changes drastically for the multi-component systems with nonzero
boundary conditions. At present, the applicability of the IST is restricted to the specific boundary conditions in which all the components have the same wavenumber 
and frequency at infinity. See  [20, 21], for instance.
We recall that there exists a unified approach for constructing multisoliton solutions of integrable PDEs which is based on a method of algebraic geometry [22].
\par
 The BD transformation is a useful tool to construct exact solutions of completely integrable PDEs [4-6]. In this framework, one needs to solve the linear eigenvalue problem 
 associated with
the Lax pair with a given seed solution  of the nonlinear PDE. Once this procedure has been completed, the construction of solutions can be performed by purely algebraic mean.
Although the successive application of the BD transformation to seed solutions allows us to obtain multisoliton solutions, the algebra involved becomes formidable as the number of
components increases.
A systematic method has been developed for constructing soliton solutions of the $n$-component NLS system with plane-wave boundary conditions [23-26].  See also [27] and
references theirin. A difficulty of the method is that one must solve the
characteristic equation of the Lax pair with a given seed solution.  Consequently,  the problem reduces to find eigenvalues of the $(n+1)\times(n+1)$ constant matrices
whose entries are complex numbers. It turns out that
even in the simplest 2-component system,
the computation of eigenvalues is really troublesome, making  the derivation of one-solutions  difficult to address [23]. 
\par
The third  approach which will be employed in our analysis is  the direct method [7-9].  While both the IST and BD transformation 
are based  on the Lax representations of completely integrable PDEs, the direct method does not need the knowledge of the IST, and it can also be applicable to nonintegrable equations.
It provides a powerful tool to obtain  special solutions such as soliton and periodic solutions, even though it is not amenable to the analysis of the
general initial value problems.   The starting point of the method is  to find appropriate dependent variable transformations to bilinearize a given
system of PDEs.  The most of the existing integrable PDEs can be bilinearized and the multisoliton solutions of the corresponding bilinear equations are constructed
in systematic ways.  Now, the application of the direct method to the  system (1.1) has been done for various settings.  The most results are concerned with the multi-component NLS system.
 A number of works have been devoted to studying the physically important two-component system  which is sometimes called the Manakov system [28].  
An article [29] reviews various types of solutions for the system with focusing, defocusing and mixed type nonlinearities.  In particular,  it addresses the properties of the
one- and two soliton solutions in detail.  Quite recently, the $n$-component NLS system subjected to the boundary conditions (1.2) and (1.3) was analyzed based on the reduction method of the multi-component
KP hierarchy, and  the determinantal expressions of the $N$-soliton solutions were presented [30].  
See also [31] for the 2-component NLS system with the boundary conditions (1.2) where the dark-dark solitons and their dynamics are explored in quite some detail.
In the most general case with $\mu\not=0$ and $\gamma\not= 0$, we have constructed the bright $N$-soliton solution 
by imposing zero boundary conditions at infinity i.e., $q_j\rightarrow 0,\ |x|\rightarrow \infty \ (j=1, 2, ..., n)$. Specifically, we have presented
the two different expressions of the $N$-soliton solution both of which have a simple structure in terms of determinants.  See theorem 2.1 and theorem 5.1 in [13]. 
The specific two-component system has been considered in [32, 33].   
To the best of our knowledge, the $N$-soliton formulas for the general $n$-component system (1.1) with the boundary conditions 1) and 2)
are presented here for the first time.  They unify most of the existing soliton solutions of physically important equations of the NLS type. 
\par
The outline of this paper is the following. In section 2, we introduce a few  notations  and some basic formulas for determinants.
In section 3, we bilinearize the system (1.1) with the boundary conditions (1.2) through appropriate dependent variable transformations.
Subsequently, the construction of the $N$-soliton solution is done following the standard recipe of the direct method.
It is shown that the tau functions associated with the $N$-soliton solution have a simple structure expressed in terms of determinants, and
the proof of the $N$-soliton solution is performed by means of an elementary theory of determinants. 
Unlike the bright soliton solutions with zero boundary conditions presented in [13, 33],
certain constraints must be imposed on soliton parameters, or in the terminology of the IST, the eigenvalues of the Lax operator. 
This leads to a difficulty  in  investigating soliton solutions of the $n$-component system for large $n$.
 Actually, one must solve the algebraic equation of order $n$  with which  the real and imaginary parts of the amplitude parameter of each soliton are connected.
 We exemplify this statement for the one-soliton solutions.
 It is important that the algebraic equations arising from constraints have real coefficients, and hence they are tractable when compapred with the
 characteristic equations with complex coefficients in the BD formulation.
In section 4, the similar bilinearization is performed for the system (1.1)  under the boundary
conditions (1.3) and the explicit $N$-soliton solution is presented.  The structure of the $N$-soliton solution is found to be quite different from that given in section 3 for the pure
plane-wave background fields. To be more specific, no constraints are imposed on the soliton parameters. This important feature facilitates the analysis of  soliton solutions.
As an example, the one-soliton solutions are given explicitly and  their limiting profiles  as $\gamma$ tends to zero are discussed.
Section 5 is devoted to concluding remarks. In appendices A-C, the three lemmas are proved which play the important role in establishing the $N$-soliton formulas.
\par
\bigskip
\leftline{\bf 2. Notation and basic formulas for determinants}\par
\bigskip
\noindent First of all,  we introduce for convenience the notation and some basic formulas for determinants associated with the
$N$-soliton solutions. \par
\medskip
\noindent{2.1.} Notation  \par
\medskip
\noindent Let $D=(d_{jk})_{1\leq j ,k\leq N}$ be an $N\times N$ matrix and ${\bf a}=(a_j)_{1\leq j\leq N}, {\bf b}=(b_j)_{1\leq j\leq N}, {\bf c}=(c_j)_{1\leq j\leq N},  {\bf d}=(d_j)_{1\leq j\leq N}$        
be the $N$-component row vectors, where all the entries in the  matrices and vectors are complex-valued functions of $x$ and $t$. The bordered marices associated with the matrix $D$ are defined by
$$D({\bf a}; {\bf b})=\begin{pmatrix} D & {\bf b}^T\\ {\bf a} & 0\end{pmatrix},
\quad D({\bf a}, {\bf b } ; {\bf c}, {\bf d})=\begin{pmatrix} D & {\bf c}^T & {\bf d}^T\\ {\bf a} & 0 &0 \\ {\bf b} & 0 & 0\end{pmatrix}, \eqno(2.1)$$ 
where the symbol $T$ denotes transpose.       
The first cofactor of $d_{ij}$ is defined by $D_{ij}=\partial |D|/\partial d_{ij},\ (|D|={\rm det}\,D)$ and 
the second cofactor by $D_{ij, pq}=\partial^2|D|/\partial d_{ip}\partial d_{jq}\ (i<j, p<q)$. \par
\bigskip
\newpage
\noindent{2.2.} Basic formulas for determinants \par
\medskip
\noindent The following formulas are well-known in the theory of determinants  and  will be used frequently in our analysis [34]: 
$${\partial |D|\over \partial x}=\sum_{i,j=1}^N{\partial d_{ij}\over \partial x}\,D_{ij}, \eqno(2.2)$$
$$\begin{vmatrix} D & {\bf a}^T\\ {\bf b} & z\end{vmatrix}=|D|z-\sum_{i,j=1}^ND_{ij}a_ib_j,  \eqno(2.3)$$
$$|D({\bf a}, {\bf b}; {\bf c}, {\bf d})||D|= |D({\bf a}; {\bf c})||D({\bf b}; {\bf d})|-|D({\bf a}; {\bf d})||D({\bf b}; {\bf c})|, \eqno(2.4)$$
$$\delta_{ij}|D|=\sum_{k=1}^Nd_{ik}D_{jk}=\sum_{k=1}^Nd_{ki}D_{kj}, \eqno(2.5)$$
\begin{align}
D_{ij} &= \sum_{p=1}^Nd_{pq}\,D_{ip, jq}\quad (j\not=q) \tag{2.6a} \\
&= \sum_{q=1}^Nd_{pq}\,D_{ip, jq}\quad (i\not=p), \tag{2.6b}
\end{align}
$$|D({\bf a}, {\bf b } ; {\bf c}, {\bf d})|=\sum_{\substack{i,j,p,q=1\\ (i\not=j, p\not=q)}}^Na_pb_qc_id_jD_{ij, pq}. \eqno(2.7) $$
Formula (2.2) is the differentiation rule of the determinant, and (2.3) is the expansion formula for a bordered determinant with respect to the last row and last column. 
Formula (2.4) is Jacobi's identity.  Formula (2.5) is the expansion rule of the determinant whereas formula (2.6) 
is the expansion rule of the first cofactor of $d_{ij}$ in terms of the second cofactors, where $\delta_{jk}$ is Kronecker's delta.
Formula (2.7) is the expansion rule of the determinant with double borders in terms of the second cofactors. 
\par
\bigskip
\leftline{\bf 3. The modified NLS system  with  plane-wave  boundary conditions }\par
\bigskip
\noindent{3.1. Gauge transformation}\par
\medskip
 \noindent     We consider the system of PDEs (1.1) subjected to  the plane-wave boundary conditions (1.2). 
 As in the case of  zero boundary conditions considered in [13], we first apply the following gauge transformation 
 $$q_j=u_j\,{\rm exp}\left[-{{\rm i}\gamma\over 2}\int_{-\infty}^x\sum_{s=1}^n\sigma_s(|u_s|^2-\rho_s^2)dx\right], \quad  j=1, 2, ..., n, \eqno(3.1)$$
 to the system, where $u_j=u_j(x,t)\  (j=1, 2, ..., n)$ are complex-valued functions of $x$ and $t$.
 The system (1.1) is then transformed  to the system of nonlinear PDEs for $u_j$:
 $${\rm i}u_{j,t}+u_{j,xx}+{\rm i}\lambda\gamma u_{j,x}+{\rm i}\gamma\left(\sum_{s=1}^n\sigma_su_s^*u_{s,x}-{\rm i}\sum_{s=1}^n\sigma_sk_s\rho_s^2\right)u_j$$
 $$+\left[\left(\mu-{\lambda\gamma^2\over 2}\right)\sum_{s=1}^n\sigma_s|u_s|^2+{(\lambda\gamma)^2\over2}\right]u_j=0, \quad  j=1, 2, ..., n. \eqno(3.2)$$
Here, we have put  $\lambda=\sum_{s=1}^n\sigma_s\rho_s^2$ for simplicity and the asterisk appended to $u_s$ denotes complex conjugate. \par
\bigskip
\noindent 3.2. Bilinearization \par
\medskip
\noindent The following proposition is the starting point in our analysis. \par
\medskip
\noindent {\bf Proposition 3.1.} {\it By means of the dependent variable transformations
$$u_j=\rho_j{\rm e}^{{\rm i}(k_jx-\omega_jt)}\,{h_j\over f},\quad j=1, 2, ..., n, \eqno(3.3)$$
the system of nonlinear PDEs (3.2) can be decoupled into the following system of bilinear equations for $f$ and $h_j$
$${\rm i}D_th_j\cdot f+{\rm i}(2k_j+\lambda\gamma)D_xh_j\cdot f+D_x^2h_j\cdot f=0,\quad j=1, 2, ..., n, \eqno(3.4)$$
$$D_xf\cdot f^*-{{\rm i}\gamma\over 2}\sum_{s=1}^n\sigma_s\rho_s^2(h_sh_s^*-ff^*)=0, \eqno(3.5)$$
$$D_x^2f\cdot f^*-{{\rm i}\gamma\over 2}\sum_{s=1}^n\sigma_s\rho_s^2D_xh_s\cdot h_s^*+\gamma\sum_{s=1}^n\sigma_sk_s\rho_s^2(h_sh_s^*-ff^*)$$
$$+\left({\lambda\gamma^2\over 4}-\mu\right)\sum_{s=1}^n\sigma_s\rho_s^2(h_sh_s^*-ff^*)=0. \eqno(3.6)$$
Here, $f=f(x, t)$ and $h_j=h_j(x, t)\ (j=1, 2, ..., n)$ are complex-valued functions of $x$ and $t$, and
the bilinear operators $D_x$ and $D_t$ are defined by
$$D_x^mD_t^nf\cdot g=\left({\partial\over\partial x}-{\partial\over\partial x^\prime}\right)^m
\left({\partial\over\partial t}-{\partial\over\partial t^\prime}\right)^n
f(x, t)g(x^\prime,t^\prime)\Big|_{ x^\prime=x,\,t^\prime=t},  \eqno(3.7)$$
where $m$ and $n$ are nonnegative integers.} \par 
\bigskip
\noindent{\bf Proof.} Substituting (3.3) into (3.2) and using the definition of the bilinear operators,  the sysytem (3.2) can be rewritten in the form
$${1\over f^2}\left\{{\rm i}D_th_j\cdot f+{\rm i}(2k_j+\lambda\gamma)D_xh_j\cdot f+D_x^2h_j\cdot f\right\}$$
$$+{h_j\over f^3f^*}\Biggl[-f^*D_x^2f\cdot f+{\rm i}\gamma\Big\{\sum_{s=1}^n\sigma_s\rho_s^2h_s^*D_xh_s\cdot f+{\rm i}f\sum_{s=1}^n\sigma_sk_s\rho_s^2(h_sh_s^*-ff^*)\Big\}$$
$$+\left(-{\lambda\gamma\over 2}+\mu\right)f\sum_{s=1}^n\sigma_s\rho_s^2(h_sh_s^*-ff^*)\Biggr]=0. \eqno(3.8)$$
 Inserting the identity
$$f^*D_x^2f\cdot f=fD_x^2f\cdot f^*-2f_xD_xf\cdot f^*+f(D_xf\cdot f^*)_x, \eqno(3.9)$$
into the corresponding term in (3.8), we can put it into the form
$${1\over f^2}\left\{{\rm i}D_th_j\cdot f+{\rm i}(2k_j+\lambda\gamma)D_xh_j\cdot f+D_x^2h_j\cdot f\right\}$$
$$+{h_j\over f^3f^*}\Biggl[f\Biggl\{-D_x^2f\cdot f^*-(D_xf\cdot f^*)_x+{\rm i}\gamma\Bigl\{\sum_{s=1}^n\sigma_s\rho_s^2h_s^*h_{s,x}+{\rm i}\sum_{s=1}^n\sigma_sk_s\rho_s^2(h_sh_s^*-ff^*)\Bigr\}$$
$$+\left(-{\lambda\gamma^2\over 2}+\mu\right)\sum_{s=1}^n\sigma_s\rho_s^2(h_sh_s^*-ff^*)\Biggr\}
+f_x\left(2D_xf\cdot f^*-{\rm i}\gamma\sum_{s=1}^n\sigma_s\rho_s^2h_sh_s^*\right)\Biggr]=0. \eqno(3.10)$$
One can confirm that the left-hand side of (3.10) becomes zero by using the bilinear equations (3.4)-(3.6).   \hspace{\fill}$\Box$
\bigskip
\par
The fundamental quantities $f$ and $h_j$ characterize completely  solutions. They are sometimes called the tau functions.
While the expressions (3.3) for $u_j$ give the solutions of the system of PDEs  (3.2) in terms of the tau functions $f$ and $h_j\ (j=1, 2, ..., n)$, the original variables $q_j$ are expressible by them as well.
To show this, we use (3.3) and (3.5) to obtain the relation
$${\partial\over \partial x}\,{\rm ln}\,{f\over f^*}=-{{\rm i}\gamma\over 2}\sum_{s=1}^n\sigma_s(|u_s|^2-\rho_s^2). \eqno(3.11)$$
If we introduce (3.11) into (3.1), integrate with respect to $x$ and note (3.3), we find the desired expressions
$$q_j=\rho_j{\rm e}^{{\rm i}(k_jx-\omega_jt)}\,{h_jf^*\over f^2},\quad j=1, 2, ..., n. \eqno(3.12)$$
\par
\medskip
\noindent{\bf Remark 3.1.} If $\gamma=0$, then the bilinear equation (3.5) reduces to  $D_xf\cdot f^*=0$, implying that $f^*=c(t)f$. An arbitrary function $c(t)$ can be set to 1 by 
imposing a boundary condition, $f=1, \ x\rightarrow -\infty$, for instance.  Thus, $q_j$ from (3.12) simplify to
$$q_j=\rho_j{\rm e}^{{\rm i}(k_jx-\omega_jt)}\,{h_j\over f},\quad j=1, 2, ..., n, \eqno(3.13)$$
and they satisfy the $n$-component NLS system with the plane-wave boundary conditions (1.2)
$${\rm i}\,q_{j,t}+q_{j,xx}+\mu\left(\sum_{s=1}^n\sigma_s|q_s|^2\right)q_j=0,\quad j=1, 2, ..., n. \eqno(3.14)$$
\par
\bigskip
\noindent{3.3.} The dark $N$-soliton solution \par
\medskip
\noindent Here, we establish the following theorem. \par
\medskip
\noindent{\bf Theorem 3.1.}\ {\it The N-soliton solution of the system of bilinear equations (3.4)-(3.6) is given in terms of the following determinants
$$f=|D|, \quad h_s=|H_s|, \quad s=1, 2, ..., n, \eqno(3.15a)$$
$$D=(d_{jk})_{1\leq j,k\leq N}, \quad d_{jk}=\delta_{jk}-{{\rm i}p_j+{\mu\over\gamma}\over p_j+p_k^*}\,z_jz_k^*, \eqno(3.15b)$$
$$H_s=(h_{jk}^{(s)})_{1\leq j,k\leq N}, \quad h_{jk}^{(s)}=\delta_{jk}+{\left({\rm i}p_j+{\mu\over\gamma}\right)(p_j-{\rm i}k_s)\over (p_j+p_k^*)(p_k^*+{\rm i}k_s)}\,z_jz_k^*, \eqno(3.15c)$$
$$z_j={\rm exp}[p_jx+({\rm i}p_j^2-\lambda\gamma p_j)t+\zeta_{j0}], \quad j=1, 2, ..., N, \quad \lambda=\sum_{s=1}^n\sigma_s\rho_s^2. \eqno(3.15d)$$
Here, $p_j$ and $\zeta_{j0}\ (j=1, 2, ..., N)$ are arbitrary complex parameters characterizing the amplitude and phase of the solitons, respectively,
 and the $N$ constraints are imposed on the parameters $p_j$
$${\gamma\over 2}\sum_{s=1}^n\sigma_s\rho_s^2\, {{\rm i}(p_j-p_j^*)+k_s+{\mu\over\gamma}\over (p_j-{\rm i}k_s)(p_j^*+{\rm i}k_s)}=-1, \quad j=1, 2, ..., N. \eqno(3.16)$$}
\par
\bigskip
\noindent {\bf Remark 3.2.} By means of the transformation of the variables, $f=\tilde f, h_s=\tilde h_s\ (s=1, 2, ..., n), \, x=\tilde x+{2\mu\over \gamma}\,\tilde t$ combined with
the transformation of the parameters, $p_j=\tilde p_j+{\rm i}\,{\mu\over\gamma}\ (j=1, 2, ..., N), k_s=\tilde k_s+{\mu\over\gamma}\ (s=1, 2, ..., n)$, the form of the
bilinear equations (3.4) and (3.5) is unchanged whereas the bilinear equation (3.6) reduces to a simplified form with $\mu=0$.  
The $N$-soliton solution (3.15) and the constraints  (3.16)  remain the same form with $\mu=0$. Thus, the proof of the $N$-soliton solution
may be performed under the setting $\mu=0$ without loss of generality. 
\par
\bigskip
\noindent {\bf Remark 3.3.} Unlike the bright $N$-soliton solution with zero boundary conditions [13],  the real part of $p_j$ is related to its imaginary part
by (3.16). In the general $n$-component system, one needs to solve the algebraic equation of order $n$ for $({\rm Re}\,p_j)^2$.  As well-known, analytical
solutions are not available for $n\geq 5$. If the wavenumbers $k_j\ (j=1, 2, ..., n)$ take the same value which implies that all the components of the system have the same asymptotic
form except the amplitudes $\rho_j$  and the phase constants $\phi_j^{(\pm)}$ (see (1.2)), then the constraints (3.16) reduce simply to a single quadratic equation for ${\rm Re}\,p_j$.  This special case has been dealt with
by means of the IST  for the multi-component NLS system [20, 21]. \par
\bigskip
\noindent {\bf Remark 3.4.} The multi-component NLS system (i.e., $\gamma=0$ in (1.1)) is invariant under the transformation $t=-\tilde t, x={\rm i}\tilde x, \mu=-\tilde\mu$.
Suppose that $\sigma_s=1\ (s=1, 2, ..., n)$ and $\tilde\mu>0$. We change the parameters in (3.15) according to  the rule $k_s=-{\rm i}\tilde k_s, \omega_s=-\tilde\omega_s\ (s=1, 2, ..., n)$,
 $p_j=-{\rm i}\tilde p_j, \zeta_{j0}=\tilde\zeta_{j0}+{\rm ln}\sqrt{\gamma}\ (j=1, 2, ..., N)$ and then take the limit $\gamma\rightarrow 0$.  The resulting expression  gives rise to the  $N$-soliton solution
of the focusing NLS system with plane-wave boundary conditions. The solutions thus constructed exhibit a rich mathematical structure. Specifically, a reduction procedure applied to the
 soliton solutions would produce the breather and rogue wave solutions, as already demonstrated for the  soliton solutions of the focusing NLS equation [35]. \par
\bigskip
The proof of theorem 3.1 will be performed by using  a sequence of lemmas, which we shall summarize. In accordance with remark 3.2, we put $\mu=0$ in  formulas that follow. \par
\bigskip
\noindent {\bf Lemma 3.1.} {\it The expression of $h_s$ from $(3.15a)$ is rewritten in the form
$$h_s=|D|-{{\rm i}\over k_s}\,|D({\bf z}_s^*; {\bf z}_x)|, \quad s=1, 2, ..., n, \eqno(3.17a)$$
where 
$${\bf z}=(z_j)_{1\leq j\leq N}, \quad {\bf z}_s=\left({k_s\over p_j-{\rm i}k_s}\,z_j\right)_{1\leq j\leq N},\quad s=1, 2, ..., n, \eqno(3.17b)$$
 are $N$-component row vectors.} \par  
\bigskip
\noindent {\bf Proof.} Using an identity
$${{\rm i}p_j\over p_j+p_k^*}\,{p_j-{\rm i}k_s\over p_k^*+{\rm i}k_s}=-{{\rm i}p_j\over p_j+p_k^*} +{{\rm i}p_j\over p_k^*+{\rm i}k_s},$$
the determinant $|H_s|$ from  $(3.15c)$  is modified in the form
\begin{align}
|H_s| &= \left|\left(d_{jk}+{{\rm i}p_j\over p_k^*+{\rm i}k_s}\,z_jz_k^*\right)_{1\leq j,k\leq N}\right| \notag \\
&=\begin{vmatrix} D &  {\rm i}{\bf z}_x^T\\ -{1\over k_s}\,{\bf z}_s^* & 1\end{vmatrix}. \notag 
\end{align}
Applying formula (2.3) to the last expression gives (3.17).  \hspace{\fill}$\Box$
\par
\bigskip
The following lemma provides the differentiation rules of $f$ and $h_s$ with respect to $t$ and $x$. \par
\medskip 
\noindent {\bf Lemma 3.2.} 
$$f_t={\rm i}|D({\bf z}^*; {\bf z}_t)|+|D({\bf z}^*_x; {\bf z}_x)|, \eqno(3.18)$$
$$f_x={\rm i}|D({\bf z}^*; {\bf z}_x)|, \eqno(3.19)$$
$$f_{xx}={\rm i}|D({\bf z}^*_x; {\bf z}_x)|+{\rm i}|D({\bf z}^*; {\bf z}_{xx})|, \eqno(3.20)$$
$$h_{s,t}={\rm i}|D({\bf z}^*; {\bf z}_t)|+|D({\bf z}^*_x; {\bf z}_x)|-{{\rm i}\over k_s}|D({\bf z}^*_{s,t}; {\bf z}_x)|-{{\rm i}\over k_s}|D({\bf z}^*_{s}; {\bf z}_{xt})|
+{1\over k_s}|D({\bf z}_s^*, {\bf z}^*; {\bf z}_x, {\bf z}_t)|, \eqno(3.21)$$
$$h_{s,x}=-|D({\bf z}^*_s; {\bf z}_x)|-{{\rm i}\over k_s}|D({\bf z}^*_{s}; {\bf z}_{xx})|, \eqno(3.22)$$
$$h_{s,xx}={\rm i}|D({\bf z}^*_x; {\bf z}_x)|+{\rm i}|D({\bf z}^*; {\bf z}_{xx})|-{{\rm i}\over k_s}|D({\bf z}^*_{s,xx}; {\bf z})|-{2{\rm i}\over k_s}|D({\bf z}^*_{s,x}; {\bf z}_{xx})|$$
$$-{{\rm i}\over k_s}|D({\bf z}^*_{s}; {\bf z}_{xxx})|+{1\over k_s}|D({\bf z}_s^*, {\bf z}^*; {\bf z}_{xx}, {\bf z}_x)|, \eqno(3.23)$$
\bigskip
\noindent {\bf Proof.} We prove (3.18). Applying formula (2.2) to  $f$ from (3.15) gives
\begin{align}
f_t &=\sum_{j,k=1}^ND_{jk}\{{\rm i}(p_j^2-{p_k^*}^2)-\lambda\gamma(p_j+p_k^*)\}{-{\rm i}p_j\over p_j+p_k^*}\,z_jz_k^* \notag \\
&=\sum_{j,k=1}^ND_{jk}\{-{\rm i}({\rm i}p_j^2-\lambda\gamma p_j)-p_jp_k^*\}z_jz_k^*. \notag
\end{align}
 In view of formula (2.3) and the realtions $z_{j,t}=({\rm i}p_j^2-\lambda\gamma p_j)z_j, z_{j,x}=p_jz_j$, the last expression coincides with (3.18).  
 A key feature in the above computation is that the factor $(p_j+p_k^*)^{-1}$ in the element $a_{jk}$ has been canceled after differentiation with respect to $t$.
 All the other formulas can be proved in the same way by using formulas (2.2) and (2.3 ) as well as 
some basic propeerties of determinants. \hspace{\fill}$\Box$
\par
\bigskip
\noindent {\bf Lemma 3.3.} {\it The complex conjugate expressions of $f$, $f_x$ and $h_s$ are given as follows.}
\medskip
$$f^*=|D| -{\rm i}|D({\bf z}^*; {\bf z})|, \eqno(3.24)$$
$$f_x^*= -{\rm i}|D({\bf z}^*_x; {\bf z})|, \eqno(3.25)$$
$$h_s^*=|D| -{\rm i}|D({\bf z}^*; {\bf z})|+{{\rm i}\over k_s}\,|D({\bf z}_x^*; {\bf z}_s)|+{1\over k_s}|D({\bf z}_x^*, {\bf z}^*; {\bf z}_s, {\bf z})|. \eqno(3.26)$$
\bigskip
\noindent {\bf Proof.}  Let $D^\dagger$ be the Hermitian conjugate of the matrix $D$.  
Since $|D|=|D^T|$, one can see that $f^*=|D^*|=|D^\dagger|$. It follows from $(3.15b)$ that $D^\dagger =D+{\rm i}{\bf z}^T{\bf z}^*$.  These relations lead to formula (3.24)
with the help of formula (2.3). Formulas (3.25) and (3.26) can be proved in the same way by taking the complex conjugate expression of (3.19) and  (3.17), respectively.  \hspace{\fill}$\Box$
\par
\medskip
 When one tries to show that  the $N$-soliton solution (3.15) solves the bilinear equations (3.5) and (3.6), the following lemma plays the central role together with Jacobi's identity (2.4).
 \par
\bigskip
\noindent {\bf Lemma 3.4.} \par
\medskip
{\it
$$\sum_{s=1}^n{\sigma_s\rho_s^2\over k_s^2}\,|D({\bf z}_s^*; {\bf z}_s)|=|D({\bf z}^*B; {\bf z})|, \eqno(3.27)$$
$$\sum_{s=1}^n{\sigma_s\rho_s^2\over k_s^2}\,|D({\bf z}_s^*, {\bf a}^*; {\bf b}, {\bf z}_s)|=|D({\bf z}^*B, {\bf a}^*; {\bf b}, {\bf z})|
-{\rm i}(|D({\bf a}^*; {\bf b}B)|-|D({\bf a}^*B; {\bf b})|). \eqno(3.28)$$
Here, $B$ is a diagonal matrix given by
$$B={\rm diag}\,(\beta_1, \beta_2, ..., \beta_N), \quad \beta_j=\sum_{s=1}^n{\sigma_s\rho_s^2\over (p_j-{\rm i}k_s)(p_j^*+{\rm i}k_s)}, \quad j=1, 2, ..., N, \eqno(3.29)$$
 ${\bf z}^*B=(\beta_jz_j^*)_{1\leq j\leq N}$ is the $N$-component row vector, and ${\bf a}=(a_j)_{1\leq j\leq N}$ and   ${\bf b}=(b_j)_{1\leq j\leq N}$
 are arbitrary $N$-component row vectors.} \par
\medskip
The proof of lemma 3.4 will be given in appendix A.  \par
\bigskip
\noindent 3.4. Proof of theorem 3.1 \par
\medskip
\noindent Here, we show that the tau functions (3.15) associated with the dark $N$-soliton solution  solve the bilinear equations (3.4)-(3.6). The proof will be performed by employing
lemmas 3.1-3.4 and some basic formulas for determinants. In particular, Jacobi's identity (2.4) plays the central role, as in the case of the proof
of the bright $N$-soliton solution of the  modified NLS system (1.1) [13]. \par
\medskip
\noindent 3.4.1. {\bf Proof of (3.4).}  Let $P$ be the left-hand side of (3.4).  We write it in the form
$P=P_1f-P_2h_s-2h_{s,x}f_x$ with
$$P_1={\rm i}h_{s,t}+{\rm i}(2k_s+\lambda\gamma)h_{s,x}+h_{s,xx},\quad P_2={\rm i}f_t+{\rm i}(2k_s+\lambda\gamma)f_x-f_{xx}.$$ 
\par 
First, we compute $P_1$ by using (3.21)-(3.23) to obtain
$$P_1=-|D({\bf z}^*; {\bf z}_t)|+{1\over k_s}|D({\bf z}_{s,t}^*-{\rm i}(2k_s+\lambda\gamma)k_s{\bf z}_s^*-k_s{\bf z}_{s,x}^*; {\bf z}_x)|$$
$$+{1\over k_s}|D({\bf z}_s^*; {\bf z}_{xt}+(k_s+\lambda\gamma){\bf z}_{s,xx}-{\rm i}{\bf z}_{s,xxx})|
+{\rm i}|D({\bf z}^*_x; {\bf z}_x)|-{{\rm i}\over k_s}|D({\bf z}^*_{s,x}; {\bf z}_{xx})|$$
$$+{{\rm i}\over k_s}|D({\bf z}_s^*, {\bf z}^*; {\bf z}_x, {\bf z}_t)|+{1\over k_s}|D({\bf z}_s^*, {\bf z}^*; {\bf z}_{xx}, {\bf z}_x)|.$$
It follows from $(3.15d)$ and $(3.17b)$ that
$${\bf z}_{s,t}^*-{\rm i}(2k_s+\lambda\gamma)k_s{\bf z}_s^*-k_s{\bf z}_{s,x}^*=-{\rm i}k_s{\bf z}_x^*-(2k_s+\lambda\gamma)k_s{\bf z}^*,\quad 
{\bf z}_{xt}+(k_s+\lambda\gamma){\bf z}_{s,xx}-{\rm i}{\bf z}_{s,xxx}=k_s{\bf z}_{xx}, $$
which reduce  $P_1$ to 
$$P_1=-|D({\bf z}^*; {\bf z}_t+(2k_s+\lambda\gamma){\bf z}_x+{\rm i}{\bf z}_{xx})|
+{{\rm i}\over k_s}|D({\bf z}_s^*, {\bf z}^*; {\bf z}_x, {\bf z}_t)|+{1\over k_s}|D({\bf z}_s^*, {\bf z}^*; {\bf z}_{xx}, {\bf z}_x)|.$$
To compute $P_2$, we use (3.18)-(3.20), giving
$$P_2=-|D({\bf z}^*; {\bf z}_t+(2k_s+\lambda\gamma){\bf z}_x+{\rm i}{\bf z}_{xx})|.$$
\par
Last, substituting above expressions for $P_1$ and $P_2$ into $P$ and using (3.15) and (3.22) as well as the relations
$${\bf z}_t={\rm i}{\bf z}_{xx}-\lambda\gamma{\bf z}_x, \quad {\bf z}_t+(2k_s+\lambda\gamma){\bf z}_x+{\rm i}{\bf z}_{xx}=2{\rm i}{\bf z}_{xx}+2k_s{\bf z}_x,$$
 one arrives at the expression
$$P=-{2\over k_s}\Bigl[|D({\bf z}_s^*, {\bf z}^*; {\bf z}_{x}, {\bf z}_{xx})||D|-|D({\bf z}^*_{s}; {\bf z}_{x})||D({\bf z}^*; {\bf z}_{xx})|
+|D({\bf z}^*_{s}; {\bf z}_{xx})||D({\bf z}^*; {\bf z}_{x})|\Bigr].$$
In view of Jacobi's identity $(2.4)$, $P$ becomes zero.  \hspace{\fill}$\Box$ \par
\bigskip
\noindent  3.4.2. {\bf Proof of (3.5).}  Let $Q$ be the left-hand side of (3.5). Substituting (3.17), (3.19), (3.24), (3.25) and (3.26) into $Q$ and using the relation
$\sum_{s=1}^n(\sigma_s\rho_s^2/ k_s)\,{\bf z}_s=-(2{\rm i}/ \gamma)\,{\bf z}+{\bf z}_xB$
which can be derived  from  $(3.16), (3.17b)$ and (3.29), we obtain
$$Q=-{{\rm i}\gamma\over 2}\Bigl[{\rm i}|D|\{-|D({\bf z}_x^*B; {\bf z}_x)|+|D({\bf z}_x^*; {\bf z}_xB)|-{\rm i}|D({\bf z}_x^*, {\bf z}^*; {\bf z}_xB, {\bf z})|\}
-|D({\bf z}_x^*B; {\bf z}_x)||D({\bf z}^*; {\bf z})|\Bigr]$$
$$-{{\rm i}\gamma\over 2}\sum_{s=1}^n{\sigma_s\rho_s^2\over k_s^2}\Bigl[|D({\bf z}^*_s; {\bf z}_x)||D({\bf z}^*_x; {\bf z}_s)|-{\rm i}|D({\bf z}^*_s; {\bf z}_x)||D({\bf z}_x^*, {\bf z}^*; {\bf z}_s, {\bf z})|\Bigr].$$
\par
It follows from Jacobi's identity $(2.4)$ that
$$|D({\bf z}_x^*, {\bf z}^*; {\bf z}_xB, {\bf z})||D|=|D({\bf z}_x^*; {\bf z}_xB)||D({\bf z}^*; {\bf z})|-|D({\bf z}_x^*; {\bf z})||D({\bf z}^*; {\bf z}_xB)|,$$
$$|D({\bf z}^*_s; {\bf z}_x)||D({\bf z}^*_x; {\bf z}_s)|=|D({\bf z}_s^*, {\bf z}_x^*; {\bf z}_x, {\bf z}_s)||D|+|D({\bf z}^*_s; {\bf z}_s)||D({\bf z}_x^*; {\bf z}_x)|, $$
$|D({\bf z}^*_s; {\bf z}_x)||D({\bf z}_x^*, {\bf z}^*; {\bf z}_s, {\bf z})|$
$$=|D({\bf z}_s^*, {\bf z}_x^*; {\bf z}_x, {\bf z}_s)||D({\bf z}^*; {\bf z})|
-|D({\bf z}_s^*, {\bf z}^*; {\bf z}_x, {\bf z}_s)||D({\bf z}^*_x; {\bf z})|+|D({\bf z}_x^*, {\bf z}^*; {\bf z}_x, {\bf z})||D({\bf z}^*_s; {\bf z}_s)|. $$
The third formula may be proved by multiplying $|D|$ on both sides and using Jacobi's identity (2.4). If we identify ${\bf a}={\bf z}_x, {\bf b}={\bf z}_x$ and 
${\bf a}={\bf z}, {\bf b}={\bf z}_x$, respectively in (3.28), then we obtain
$$\sum_{s=1}^n{\sigma_s\rho_s^2\over k_s^2}|D({\bf z}_s^*, {\bf z}_x^*; {\bf z}_x, {\bf z}_s)|=|D({\bf z}^*B, {\bf z}_x^*; {\bf z}_x, {\bf z})|
-{\rm i}(|D({\bf z}_x^*; {\bf z}_xB)|-|D({\bf z}_x^*B; {\bf z}_x)|),$$
$$\sum_{s=1}^n{\sigma_s\rho_s^2\over k_s^2}|D({\bf z}_s^*, {\bf z}^*; {\bf z}_x, {\bf z}_s)|=|D({\bf z}^*B, {\bf z}^*; {\bf z}_x, {\bf z})|
-{\rm i}(|D({\bf z}^*; {\bf z}_xB)|-|D({\bf z}^*B; {\bf z}_x)|).$$
\par
The last step is to substitute (3.27) and the above five relations into $Q$, which recasts it to
$$Q={\gamma\over 2}\Bigl[-|D({\bf z}^*; {\bf z})||D({\bf z}^*B, {\bf z}_x^*; {\bf z}_x, {\bf z})|+|D({\bf z}^*_x; {\bf z})||D({\bf z}^*B, {\bf z}^*; {\bf z}_x, {\bf z})|$$
$$-|D({\bf z}^*B; {\bf z})||D({\bf z}_x^*, {\bf z}^*; {\bf z}_x, {\bf z})|\Bigr]. $$
This expression becomes zero by Jacobi's identity $(2.4)$. \hspace{\fill}$\Box$ \par
\bigskip
\noindent 3.4.3. {\bf Proof of (3.6).}  We replace the last term of the left-hand side of (3.6) by $D_xf\cdot f$ from (3.5) and then  add  $Q_x(=0))$ to the resulting expression, and
show that $R\equiv 2R_1-{\rm i}\gamma R_2+\gamma R_3=0$, where
$$R_1=f_{xx}f^*-f_xf_x^*, \quad R_2=\sum_{n=1}^n\sigma_s\rho_s^2(h_{s,x}h_s^*-ff_x^*), \quad R_3=\sum_{n=1}^n\sigma_sk_s\rho_s^2(h_sh_s^*-ff^*).$$
\par
Now, substituting (3.17), (3.19), (3.20), (3.24) and (3.25) into $R_1$, we obtain
$$R_1={\rm i}(|D({\bf z}_x^*; {\bf z}_x)|+|D({\bf z}_x^*; {\bf z}_{xx})|)f^*-|D({\bf z}^*; {\bf z}_x)||D({\bf z}_x^*; {\bf z})|.$$
\par
The expression of $R_2$ can be recast, after using (3.22), (3.25) and (3.26) as well as the relation
$\sum_{s=1}^n(\sigma_s\rho_s^2/ k_s)\,{\bf z}_s^*=(2{\rm i}/ \gamma)\,{\bf z}^*+{\bf z}_x^*B$,
to 
$$R_2=-\left\{\sum_{s=1}^n\sigma_s\rho_s^2|D({\bf z}_s^*; {\bf z}_x)|-{2\over\gamma}|D({\bf z}^*; {\bf z}_{xx})|+{\rm i}|D({\bf z}_x^*B; {\bf z}_{xx})|\right\}f^*+{\rm i}\lambda|D||D({\bf z}_x^*; {\bf z})|$$
$$-\sum_{s=1}^n{\sigma_s\rho_s^2\over k_s}\Bigl[{\rm i}|D({\bf z}_s^*; {\bf z}_x)||D({\bf z}_x^*; {\bf z}_s)|-{1\over k_s}|D({\bf z}_s^*; {\bf z}_{xx})||D({\bf z}_x^*; {\bf z}_s)|$$
$$+|D({\bf z}_s^*; {\bf z}_x)||D({\bf z}_x^*, {\bf z}^*; {\bf z}_s, {\bf z})|+{{\rm i}\over k_s}|D({\bf z}_s^*; {\bf z}_{xx})||D({\bf z}_x^*, {\bf z}^*; {\bf z}_s, {\bf z})|\Bigr].$$
\par
It follows from (3.17), (3.24) and (3.26) that
$$R_3=|D|\sum_{s=1}^n\sigma_s\rho_s^2({\rm i}|D({\bf z}_x^*; {\bf z}_s)|+|D({\bf z}_x^*, {\bf z}^*; {\bf z}_s, {\bf z})|)-{\rm i}\lambda |D({\bf z}_s^*; {\bf z}_x)|f^*$$
$$-{\rm i}\sum_{s=1}^n{\sigma_s\rho_s^2\over k_s}|D({\bf z}_s^*; {\bf z}_x)|({\rm i}|D({\bf z}_x^*; {\bf z}_s)|+|D({\bf z}_x^*, {\bf z}^*; {\bf z}_s, {\bf z})|).$$
Referring to the above expressions of $R_1, R_2$ and $R_3$, $R$ becomes
$$R=2{\rm i}|D({\bf z}_x^*; {\bf z}_x)|f^*-2|D({\bf z}^*; {\bf z}_x)||D({\bf z}_x^*; {\bf z})|-{\rm i}\gamma\Bigl[-{\rm i}|D({\bf z}_x^*B; {\bf z}_{xx})|f^*+{\rm i}\lambda |D||D({\bf z}_x^*; {\bf z})|$$
$$-\sum_{s=1}^n{\sigma_s\rho_s^2\over k_s^2}|D({\bf z}_s^*; {\bf z}_x)|\{-|D({\bf z}_x^*; {\bf z}_s)|+{\rm i}|D({\bf z}_x^*, {\bf z}^*; {\bf z}_s, {\bf z})|\}\Bigr]$$
$$+\gamma |D|\sum_{s=1}^n\sigma_s\rho_s^2({\rm i}|D({\bf z}_x^*; {\bf z}_s)|+|D({\bf z}_x^*, {\bf z}^*; {\bf z}_s, {\bf z})|).$$
To simplify $R$ further, we use the relation $\sum_{s=1}^n\sigma_s\rho_s^2\,{\bf z}_s={\rm i}\lambda {\bf z}-(2/ \gamma)\,{\bf z}_x-{\rm i}{\bf z}_{xx}B$ in the last term
 and (3.24) for $f^*$, and then
apply Jacobi's identity (2.4), resulting in
$$R=-{\rm i}\gamma\Bigl[-{\rm i}|D({\bf z}_x^*B; {\bf z}_{xx})|(|D|-{\rm i}|D({\bf z}_x^*; {\bf z}_s)|)$$
$$+\sum_{s=1}^n{\sigma_s\rho_s^2\over k_s^2}|D({\bf z}_s^*; {\bf z}_{xx})|\{|D({\bf z}_x^*; {\bf z}_s)|-{\rm i}|D({\bf z}_x^*, {\bf z}^*; {\bf z}_s, {\bf z})|\}\Bigr]$$
$$+\gamma |D|\Bigl[|D({\bf z}_x^*; {\bf z}_{xx}B)|-{\rm i}|D({\bf z}_x^*, {\bf z}^*; {\bf z}_{xx}B, {\bf z})|\Bigr].$$
\par 
An application of  Jacobi's identity (2.4) gives 
$$|D({\bf z}^*_s; {\bf z}_{xx})||D({\bf z}_x^*, {\bf z}^*; {\bf z}_s, {\bf z})|$$
$$=|D({\bf z}_s^*, {\bf z}_x^*; {\bf z}_{xx}, {\bf z}_s)||D({\bf z}^*; {\bf z})|
-|D({\bf z}_s^*, {\bf z}^*; {\bf z}_{xx}, {\bf z}_s)||D({\bf z}^*_x; {\bf z})|+|D({\bf z}_x^*, {\bf z}^*; {\bf z}_{xx}, {\bf z})||D({\bf z}^*_s; {\bf z}_s)|, $$
and formula (3.28) leads to
$$\sum_{s=1}^n{\sigma_s\rho_s^2\over k_s^2}|D({\bf z}_s^*, {\bf z}_x^*; {\bf z}_{xx}, {\bf z}_s)|=|D({\bf z}^*B, {\bf z}_x^*; {\bf z}_{xx}, {\bf z})|
-{\rm i}(|D({\bf z}_x^*; {\bf z}_xB)|-|D({\bf z}_x^*B; {\bf z}_{xx})|),$$
$$\sum_{s=1}^n{\sigma_s\rho_s^2\over k_s^2}|D({\bf z}_s^*, {\bf z}^*; {\bf z}_{xx}, {\bf z}_s)|=|D({\bf z}^*B, {\bf z}^*; {\bf z}_{xx}, {\bf z})|
-{\rm i}(|D({\bf z}^*; {\bf z}_{xx}B)|-|D({\bf z}^*B; {\bf z}_{xx})|).$$
Using these relations together with (3.27), one obtains
$$\sum_{s=1}^n{\sigma_s\rho_s^2\over k_s^2}|D({\bf z}_s^*; {\bf z}_{xx})||D({\bf z}_x^*, {\bf z}^*; {\bf z}_s, {\bf z})|$$
$$=-{\rm i}|D({\bf z}^*; {\bf z})|(|D({\bf z}_x^*; {\bf z}_{xx}B)|-|D({\bf z}_x^*B; {\bf z}_{xx})|)
+{\rm i}|D({\bf z}^*_x; {\bf z})|(|D({\bf z}^*; {\bf z}_{xx}B)|-|D({\bf z}^*B; {\bf z}_{xx})|).$$
Similarly,
$$\sum_{s=1}^n{\sigma_s\rho_s^2\over k_s^2}|D({\bf z}_s^*; {\bf z}_{xx})|\{|D({\bf z}_x^*; {\bf z}_s)|$$
$$=|D|\{|D({\bf z}^*B, {\bf z}^*_x; {\bf z}_{xx}, {\bf z})|
-{\rm i}(|D({\bf z}^*_x; {\bf z}_{xx}B)|-|D({\bf z}^*_xB; {\bf z}_{xx})|)\}+|D({\bf z}^*B; {\bf z})||D({\bf z}_x^*; {\bf z}_{xx}). $$
Last, substituting above two relations into $R$, one can see that $R$ becomes zero.    \hspace{\fill}$\Box$ \par
\bigskip
\noindent {\bf Remark 3.5.} We remark that the constraints (3.16) has not been used for the proof of (3.4). On the other hand, in establishing (3.5) and (3.6), one has relied on lemma (3.4)
which is essentially based on  the constraints. \par
\bigskip
\noindent 3.5. One-soliton solutions \par
\medskip
\noindent  The soliton solutions are characterized completely by the tau functions $f$ and $h_j$   given by (3.15).  Here, we
describe the feature of one-soliton solutions. The general $N$-soliton solutions will be considered elsewhere. 
 To simplify the notation, we first put $p_1=a+{\rm i}b, \ \zeta_{10}=a(x_0+{\rm i}y_0)$ \ ($a, b, x_0, y_0 \in \mathbb{R})$, and
$${b-{\mu\over\gamma}-{\rm i}a\over 2a}=\beta\,{\rm e}^{2{\rm i}\phi}\ (\beta>0),\quad {a+{\rm i}(b-k_j)\over a-{\rm i}(b-k_j)}={\rm e}^{2{\rm i}\theta_j}, \ j=1, 2, ..., n. \eqno(3.30)$$
Then, the tau functions for the one-soliton solutions  are  written compactly in the form
$$f=1+{\rm e}^{2a(\xi+\xi_0)+2{\rm i}\phi}, \quad h_j=1-{\rm e}^{2a(\xi+\xi_0)+2{\rm i}(\theta_j+\phi)},\ j=1, 2, ..., n, \eqno(3.31)$$
with $\xi=x-(2b+\lambda\gamma)t+x_0$, and $\xi_0=(1/2a){\rm ln}\, \beta$. Formula (3.12) with (3.31) gives one-soliton solutions. 
The square modulas of the complex variable  $q_j$ is computed as
$$|q_j|^2=\rho_j^2-{2a^2\,{\rm sgn}\,a\over \sqrt{a^2+(b-\mu/\gamma)^2}}\,{\rho_j^2(2b-\mu/\gamma-k_j)\over a^2+(b-k_j)^2}
{1\over \cosh 2a(\xi+\xi_0)+{(b-\mu/\gamma){\rm sgn}\,a\over \sqrt{a^2+(b-\mu/\gamma)^2}}}. \eqno(3.32)$$
\par
In the simplest one-soliton case, the constraints (3.16) reduce simply to a single relation which connects $a$ with $b$
$${\gamma\over 2}\sum_{s=1}^n\sigma_s\rho_s^2\,{-2b+k_s+{\mu\over\gamma}\over a^2+(b-k_s)^2}=-1.  \eqno(3.33)$$
For the $n$-component system, one must solve the algebraic equation of order $n$ for $a^2$ to express $a$ in terms of $b$. 
Note, however, that  analytical solutions are obtainable up to $n=4$.  This is the main difficulty in constructing soliton solutions.
Although one can deal with the degenerate case for which all the wave numbers $k_s$ have the same (possibly zero) value, the resulting
one soliton solutions coincide essentially with those of the one-component system. \par
Last, we consider the soliton solutions of the 1-component system. Using (3.33) with $n=1$, the expression (3.32) simplifies to
$$|q_1|^2=\rho_1^2-{4a^2\,{\rm sgn}\,a\over \sigma_1\gamma\sqrt{a^2+(b-\mu/\gamma)^2}}
{1\over \cosh 2a(\xi+\xi_0)+{(b-\mu/\gamma){\rm sgn}\,a\over \sqrt{a^2+(b-\mu/\gamma)^2}}}. \eqno(3.34)$$
When the condition ${\rm sgn}(\sigma_1 \gamma a)>0$  is satisfied, then this represents the dark soliton solution with a constant background.
Taking the limit $\gamma\rightarrow 0$ under the conditions ${\rm sgn}(\sigma_1\gamma a)>0, {\rm sgn}(\gamma\mu a)<0$, the expression (3.34) reduces to 
the dark soliton solution of the NLS equation
$$|q_1|^2=\rho_1^2-{2a^2\over |\mu|}{\rm sech}^2a(\xi+\xi_0), \eqno(3.35)$$
with a constraint $a^2+(b-k_1)^2=-(\mu/2)\sigma_1\rho_1^2$ imposed on the parameters $a$ and $b$.
\par
\bigskip
\leftline{\bf 4. The modified NLS system  with  mixed type boundary conditions}\par
\bigskip
\noindent In this section, we present the $N$-soliton solution of the system of equations (1.1) with   the mixed zero and plane-wave boundary conditions (1.3).
We will see that  solutions take the form of bright-dark type solitons.
 Since the procedure for constructing the solution almost parallels  that developed for the plane-wave boundary conditions, we outline the results except for the
 proof of the $N$-soliton solution. \par
 \bigskip
 \newpage
\noindent{4.1. Gauge transformation}\par
\medskip
\noindent We first apply the  gauge transformation 
 $$q_j=u_j\,{\rm exp}\left[-{{\rm i}\gamma\over 2}\int_{-\infty}^x\left\{\sum_{s=1}^n\sigma_s|u_s|^2-\sum_{s=1}^{n-m}\sigma_{m+s}\rho_s^2\right\}dx\right], \quad  j=1, 2, ..., n, \eqno(4.1)$$
 to the system (1.1) and obtain  the system of nonlinear PDEs for $u_j$:
 $${\rm i}u_{j,t}+u_{j,xx}+{\rm i}\hat\lambda\gamma u_{j,x}+{\rm i}\gamma\left(\sum_{s=1}^n\sigma_su_s^*u_{s,x}-{\rm i}\sum_{s=1}^{n-m}\sigma_{m+s}k_s\rho_s^2\right)u_j$$
 $$+\left[\left(\mu-{\hat\lambda\gamma^2\over 2}\right)\sum_{s=1}^n\sigma_s|u_s|^2+{(\hat\lambda\gamma)^2\over2}\right]u_j=0, \quad  j=1, 2, ..., n, \eqno(4.2)$$
where  $\hat\lambda=\sum_{s=1}^{n-m}\sigma_{m+s}\rho_s^2$.  \par
\bigskip
\noindent{4.2. Bilinearization}\par
\medskip
\noindent The bilinearization of the system of nonlinear PDEs (4.2) is accomplished by the following proposition. \par
\medskip
\noindent {\bf Proposition 4.1.} {\it By means of the dependent variable transformations
$$u_j={\rm e}^{{\rm i}\hat\lambda\mu t}\,{g_j\over f}, \quad j=1, 2, ..., m, \eqno(4.3a)$$
$$u_{m+j}=\rho_j{\rm e}^{{\rm i}(k_jx-\omega_jt)}\,{h_j\over f},\quad j=1, 2, ..., n-m, \eqno(4.3b)$$
the system of nonlinear PDEs (4.2) can be decoupled into the following system of bilinear equations for $f, g_j$ and $h_j$
$${\rm i}D_tg_j\cdot f+{\rm i}\hat\lambda\gamma D_xg_j\cdot f+D_x^2g_j\cdot f=0,\quad j=1, 2, ..., m, \eqno(4.4)$$
$${\rm i}D_th_j\cdot f+{\rm i}(2k_j+\hat\lambda\gamma)D_xh_j\cdot f+D_x^2h_j\cdot f=0,\quad j=1, 2, ..., n-m, \eqno(4.5)$$
$$D_xf\cdot f^*-{{\rm i}\gamma\over 2}\left\{\sum_{s=1}^m\sigma_sg_sg_s^*+\sum_{s=1}^{n-m}\sigma_{m+s}\rho_s^2(h_sh_s^*-ff^*)\right\}=0, \eqno(4.6)$$
$$D_x^2f\cdot f^*-{{\rm i}\gamma\over 2}\left(\sum_{s=1}^m\sigma_sD_xg_s\cdot g_s^*+\sum_{s=1}^{n-m}\sigma_{m+s}\rho_s^2D_xh_s\cdot h_s^*\right)
+\gamma\sum_{s=1}^{n-m}\sigma_{m+s}k_s\rho_s^2(h_sh_s^*-ff^*)$$
$$+\left({\hat\lambda\gamma^2\over 4}-\mu\right)\left\{\sum_{s=1}^m\sigma_sg_sg_s^*+\sum_{s=1}^{n-m}\sigma_{m+s}\rho_s^2(h_sh_s^*-ff^*)\right\}=0. \eqno(4.7)$$
Here, $f=f(x, t), g_j(x, t)$ and $h_j=h_j(x, t)$ are complex-valued functions of $x$ and $t$.} \par
\bigskip
Using (4.1) and (4.6), we can express $q_j$ in terms of the tau functions $f, g_j$ and $h_j$ as
$$q_j={\rm e}^{{\rm i}\hat\lambda\mu t}\,{g_jf^*\over f^2}, \quad j=1, 2, ..., m, \eqno(4.8a)$$
$$q_{m+j}=\rho_j{\rm e}^{{\rm i}(k_jx-\omega_jt)}\,{h_jf^*\over f^2},\quad j=1, 2, ..., n-m. \eqno(4.8b)$$
\bigskip
\noindent {\bf Remark 4.1.} In the case of $\gamma=0$, the expressions (4.8) become
$$q_j={\rm e}^{{\rm i}\hat\lambda\mu t}\,{g_j\over f}, \quad j=1, 2, ..., m, \eqno(4.9a)$$
$$q_{m+j}=\rho_j{\rm e}^{{\rm i}(k_jx-\omega_jt)}\,{h_j\over f},\quad j=1, 2, ..., n-m, \eqno(4.9b)$$
and they satisfy the $n$-component NLS system (3.14). \par
\bigskip
\noindent 4.3. The bright-dark $N$-soliton solution \par
\medskip
\noindent The main result in this section is provided by the following theorem. \par
\bigskip
\noindent {\bf Theorem 4.1.} {\it The $N$-soliton solution of the system of bilinear equations (4.4)-(4.7) is given in terms of the
determinants
$$f=|D|, \quad g_s=-|D({\bf a}^*_s; {\bf z})|, \ s=1, 2, ..., m, \quad h_s=|D|+{1\over k_s}|D({\bf z}_s^*; {\bf z})|, \ s=1, 2, ..., n-m, \eqno(4.10a)$$
$$D=(d_{jk})_{1\leq j,k\leq N}, \quad d_{jk}={z_jz_k^*+{1\over2}(\mu-{\rm i}\gamma p_k^*)c_{jk}\over p_j+p_k^*}, \eqno(4.10b)$$
$$z_j={\rm exp}[p_jx+({\rm i}p_j^2-\hat\lambda\gamma p_j)t], \quad j=1, 2, ..., N, \eqno(4.10c)$$
$$c_{jk}={\sum_{s=1}^m\sigma_s\alpha_{sj}\alpha_{sk}^*\over 
1+{\gamma\over 2}\sum_{s=1}^{n-m}\sigma_{m+s}\rho_s^2\,{{\rm i}(p_j-p_k^*)+k_s+{\mu\over\gamma}\over (p_j-{\rm i}k_s)(p_k^*+{\rm i}k_s)}}, 
\quad j,k=1, 2, ..., N, \eqno(4.10d)$$
where ${\bf z}$ and ${\bf z}_s$ are $N$-component row vectors defined by $(3.17b)$ with $z_j$ given by (4.10c) and 
$${\bf a}_s=(\alpha_{s1}, \alpha_{s2}, ..., \alpha_{sN}),\quad  s=1, 2, ..., m, \eqno(4.10e)$$
are row vectors with elements $\alpha_{sj}\in \mathbb{C} \ (s=1, 2, .., m; j=1, 2, .., N)$.}
\par
\bigskip
The $N$-soliton solution is characterized by the $N$ complex parameters $p_j \ (j=1, 2, ..., N)$ and $mN$ complex parameters
$\alpha_{sj}\ (s=1, 2, ..., m; j=1, 2, ..., N)$.  The former parameters  determine the amplitude of solitons and the latter ones determine
the polarization and the envelope phases of  solitons.  
Note that we have used the same symbol as that appears in theorem 3.1 for the tau functions $f$ and $h_s$ since no
confusion would be likely to arise from this convention. 
\par
The proof of theorem 4.1 can be performed by using the differentiation rules for the tau functions $f$, $g_s$ and $h_s$ 
as well as the basic properties of determinants.
 Remark 3.2 is applied to
the current problem, so that we can put $\mu=0$ without loss of generality.  First, we summarize the differentiation rules
corresponding to those given by lemma 3.2 and lemma 3.3. \par
\bigskip
\noindent {\bf Lemma 4.1.} 
$$f_t=-{\rm i}|D({\bf z}^*; {\bf z}_x)|+{\rm i}|D({\bf z}^*_x; {\bf z})|+\hat\lambda\gamma|D({\bf z}^*; {\bf z})| , \eqno(4.11)$$
$$f_x=-|D({\bf z}^*; {\bf z})|, \eqno(4.12)$$
$$f_{xx}=-|D({\bf z}^*_x; {\bf z})|-|D({\bf z}^*; {\bf z}_{x})|, \eqno(4.13)$$
$$g_{s,t}=-|D({\bf a}_s^*; {\bf z}_t)|+{\rm i}|D({\bf a}_s^*, {\bf z}^*; {\bf z}, {\bf z}_x)|, \eqno(4.14)$$
$$g_{s,x}=-|D({\bf a}_s^*; {\bf z}_x)|, \eqno(4.15)$$
$$g_{s,xx}=-|D({\bf a}_s^*; {\bf z}_{xx})|+|D({\bf a}_s^*, {\bf z}^*; {\bf z}_x, {\bf z})|, \eqno(4.16)$$
$$h_{s,t}=-{\rm i}|D({\bf z}^*; {\bf z}_x)|+{\rm i}|D({\bf z}^*_x; {\bf z})|+\hat\lambda\gamma|D({\bf z}^*; {\bf z})|+{1\over k_s}|D({\bf z}^*_{s,t}; {\bf z})|
+{1\over k_s}|D({\bf z}^*_{s}; {\bf z}_t)|$$
$$-{{\rm i}\over k_s}|D({\bf z}_s^*, {\bf z}^*; {\bf z}, {\bf z}_x)|, \eqno(4.17)$$
$$h_{s,x}=-{\rm i}|D({\bf z}^*_s; {\bf z})|+{1\over k_s}|D({\bf z}^*_{s}; {\bf z}_x)|, \eqno(4.18)$$
$$h_{s,xx}=-{\rm i}|D({\bf z}^*_{s,x}; {\bf z})|-{\rm i}|D({\bf z}^*_s; {\bf z}_x)|+{1\over k_s}|D({\bf z}^*_{s,x}; {\bf z}_x)|
+{1\over k_s}|D({\bf z}^*_{s}; {\bf z}_{xx})|-{1\over k_s}|D({\bf z}_s^*, {\bf z}^*; {\bf z}_x, {\bf z})|. \eqno(4.19)$$
\par
\bigskip
The above formulas can be derived by using (2.2) and some basic properties of determinants.  See, for example, proof of (3.18). \par
The lemma 4.2 below provides the complex conjugate expressions of $f, f_x, g_s$ and $h_s$. \par
\bigskip
\noindent {\bf Lemma 4.2.} \par
$$f^*=|\bar D|=|(\bar d_{jk})_{1\leq j,k\leq N}|, \quad \bar d_{jk}=d_{jk}+{{\rm i}\gamma\over 2}\,c_{jk}, \eqno(4.20)$$
$$f_x^*=-|\bar D({\bf z}^*; {\bf z})|, \eqno(4.21)$$
$$g_s^*=-|\bar D({\bf z}^*; {\bf a}_s)|, \eqno(4.22)$$
$$h_s^*=|\bar D|+{1\over k_s}|\bar D({\bf z}^*; {\bf z}_s)|. \eqno(4.23)$$
\bigskip
\noindent {\bf Proof.} It follows from a well-known property of the determinant that $f^*=|D^*|=|D^\dagger|=|(d_{kj}^*)_{1\leq j,k\leq N}|$.
Referring to $(4.10b)$, one obtains $ d_{kj}^*={z_jz_k^*+{1\over2}(\mu+{\rm i}\gamma p_j)c_{kj}^*\over p_j+p_k^*}$.
Since $c_{kj}^*=c_{jk}$ by $(4.10d)$, this expression is written as $d_{kj}^*=d_{jk}+({\rm i}\gamma/2)c_{jk}=\bar d_{jk}$, which gives (4.20).
The remaining relations can be proved in the same way. \hspace{\fill}$\Box$ \par
\bigskip
The following two lemmas will be used effectively in the proof of (4.6) and (4.7). \par
\bigskip
\noindent {\bf Lemma 4.3.} \par
$$\sum_{s=1}^{n-m}{\sigma_{m+s}\rho_s^2\over k_s^2}\,|D({\bf z}_s^*; {\bf z})||\bar D({\bf z}^*; {\bf z}_s)|=
-|D(\tilde {\bf z}^*; {\bf z})||\bar D|-|\bar D({\bf z}^*; \tilde {\bf z})||D|$$
$$+{2{\rm i}\over\gamma}(|D({\bf z}^*; {\bf z})||\bar D|-|\bar D({\bf z}^*; {\bf z})||D|)
-\sum_{s=1}^{m}\sigma_s|D({\bf a}_s^*; {\bf z})||\bar D({\bf z}^*; {\bf a}_s)|, \eqno(4.24)$$
$$\sum_{s=1}^{n-m}{\sigma_{m+s}\rho_s^2\over k_s^2}\,|D({\bf z}_s^*; {\bf z}_x)||\bar D({\bf z}^*; {\bf z}_s)|=
-|D(\tilde {\bf z}^*; {\bf z}_x)||\bar D|-|\bar D({\bf z}^*; \tilde {\bf z}_x)||D|$$
$$+{2{\rm i}\over\gamma}(|D({\bf z}^*; {\bf z}_x)||\bar D|-|\bar D({\bf z}^*; {\bf z}_x)||D|)
-\sum_{s=1}^{m}\sigma_s|D({\bf a}_s^*; {\bf z}_x)||\bar D({\bf z}^*; {\bf a}_s)|. \eqno(4.25)$$
Here, $\tilde{\bf z}=(\tilde z_j)_{1\leq j\leq N}$ is an $N$-component row vector with elements $\tilde z_j=\sum_{s=1}^{n-m}\sigma_{m+s}\rho_s^2{z_j\over p_j-{\rm i}k_s}$.
\par
\bigskip
\noindent {\bf Lemma 4.4.} \par
$$|D({\bf z}_x^*; {\bf z})||\bar D|+|D({\bf z}^*; {\bf z})||\bar D({\bf z}^*; {\bf z})|+|\bar D({\bf z}^*; {\bf z}_x)||D|=0. \eqno(4.26)$$
\bigskip
The proof of lemma 4.3 and lemma 4.4 will be given in appendix B and appendix C, respectively. \par
\bigskip
\noindent 4.4. Proof of theorem 4.1 \par
\bigskip
\noindent Here, we show that the tau functions (4.10) associated with the  $N$-soliton solution solve the bilinear equations (4.4)-(4.7).
A transparent proof will be presented with the aid of lemmas 4.3 and 4.4. \par
\medskip
\noindent 4.4.1. {\bf Proof of (4.4)}  Let $\hat P$ be the left-hand side of (4.4). Substituting $(4.10a)$ and(4.11)-(4.16) into $\hat P$ and rearranging terms, one obtains
$$\hat P=-|D({\bf a}_j^*; {\rm i}{\bf z}_t+{\rm i}\hat\lambda\gamma {\bf z}_x+{\bf z}_{xx})||D|
-2|D({\bf a}_j^*, {\bf z}^*; {\bf z}, {\bf z}_x)||D|$$
$$+2(|D({\bf a}_j^*; {\bf z})||D({\bf z}^*; {\bf z}_{x})|-|D({\bf a}_j^*; {\bf z}_x)||D({\bf z}^*; {\bf z})|).$$
The first term vanishes by virtue of the relation ${\rm i}{\bf z}_t+{\rm i}\hat\lambda\gamma{\bf z}_x+{\bf z}_{xx}={\bf 0}$ which comes from $(4.10c)$.
The sum of the second and third terms becomes zero by Jacobi's identity (2.4). \hspace{\fill}$\Box$ \par
\par
\bigskip
\noindent 4.4.2. {\bf Proof of (4.5)}  Let $\hat Q$ be the left-hand side of (4.5) and write it in the form $\hat Q=\hat Q_1f-\hat Q_2h_s-2h_{s,x}f_x$ with
$$\hat Q_1={\rm i}h_{s,t}+{\rm i}(2k_s+\hat\lambda\gamma)h_{s.x}+h_{s,xx},\quad \hat Q_2={\rm i}f_{t}+{\rm i}(2k_s+\hat\lambda\gamma)f_{x}-f_{xx}. $$
Using (4.17)-(4.19), $\hat Q_1$ recasts to
$$\hat Q_1=|D({\bf z}^*+({\rm i}/k_s)(k_s+\hat\lambda\gamma){\bf z}_s^*+(1/k_s){\bf z}_{s,x}^*; {\bf z}_x)|
+{1\over k_s}|D({\bf z}_s^*; {\rm i}{\bf z}_t+k_s(2k_s+\hat\lambda\gamma){\bf z}+{\bf z}_{xx})|$$
$$+|D(({\rm i}/k_s){\bf z}^*_{s,t}-{\bf z}_x^*-{\rm i}{\bf z}^*_{s,x}; {\bf z})|+{\rm i}\hat\lambda\gamma|D({\bf z}^*; {\bf z})|
+{2\over k_s}|D({\bf z}_s^*, {\bf z}^*; {\bf z}, {\bf z}_x)|.$$
It follows from the definition of ${\bf z}$ and ${\bf z}_s$ that
$${\rm i}{\bf z}_t=-{\bf z}_{xx}+{\rm i}\hat\lambda\gamma{\bf z}_x,
 \quad  {{\rm i}\over k_s}\,{\bf z}^*_{s,t}-{\bf z}_x^*-{\rm i}{\bf z}^*_{s,x}=-{\rm i}(2k_s+\hat\lambda\gamma){\bf z}^*-(2k_s+\hat\lambda\gamma){\bf z}_s^*. $$
Taking into account the above relations, $\hat Q_1$ simplifies to
$$\hat Q_1=2\left\{|D({\bf z}^*; {\bf z}_x)|-{\rm i}k_s|D({\bf z}^*; {\bf z})|+{1\over k_s}|D({\bf z}_s^*, {\bf z}^*; {\bf z}, {\bf z}_x)|\right\}.$$
On the other hand,  introducing (4.11)-(4.13) into $\hat Q_2$, one obtains
$$\hat Q_2=2\left\{|D({\bf z}^*; {\bf z}_x)|-{\rm i}k_s|D({\bf z}^*; {\bf z})|\right\}. $$
\par
Last,  after a computation using  $\hat Q_1$ and $\hat Q_2$ above as well as $(4.10a)$, (4.12) and (4.18), 
it is found that
$$\hat Q={2\over k_s}\left\{|D({\bf z}_s^*, {\bf z}^*; {\bf z}, {\bf z}_x)||D|-|D({\bf z}^*; {\bf z}_x)||D({\bf z}_s^*; {\bf z})|+|D({\bf z}_s^*; {\bf z}_x)||D({\bf z}^*; {\bf z})|\right\},$$
which becomes zero by Jacobi's identity (2.4). \hspace{\fill}$\Box$ \par
\par
\bigskip
\noindent 4.4.3. {\bf Proof of (4.6)}  Let $\hat R$ be the left-hand side of (4.6). Substituting $(4.10a)$, (4.12), (4.20), (4.22) and (4.23) into $\hat R$, one obtains
$$\hat R=-\left\{|D({\bf z}^*; {\bf z})|+{{\rm i}\gamma\over 2}\sum_{s=1}^{n-m}{\sigma_{m+s}\rho_s^2\over k_s}\,|D({\bf z}_s^*; {\bf z})|\right\}|\bar D|$$
$$+\left\{|\bar D({\bf z}^*; {\bf z})|-{{\rm i}\gamma\over 2}\sum_{s=1}^{n-m}{\sigma_{m+s}\rho_s^2\over k_s}\,|\bar D({\bf z}^*; {\bf z}_s)|\right\}|D|$$
$$-{{\rm i}\gamma\over 2}\sum_{s=1}^{n-m}{\sigma_{m+s}\rho_s^2\over k_s^2}\,|D({\bf z}_s^*; {\bf z})||\bar D({\bf z}^*; {\bf z}_s)|
-{{\rm i}\gamma\over 2}\sum_{s=1}^{n-m}\sigma_s|D({\bf a}_s^*; {\bf z})||\bar D({\bf z}^*; {\bf a}_s)|.$$
Use (4.24) for the third term and introduce the definition of $\tilde {\bf z}$ for the first and second terms.
Then, one can see that $\hat R=0$. \hspace{\fill}$\Box$ \par
\par
\bigskip
\noindent 4.4.4. {\bf Proof of (4.7)} We add $\hat R_x(=0)$ to (4.7), replace the last term of (4.7) by $D_xf\cdot f^*$ from (4.6) and prove that $\hat S=0$, where
$$\hat S=f_{xx}f^*-f_xf_x^*
-{{\rm i}\gamma\over 2}\left(\sum_{s=1}^m\sigma_sg_{s,x}g_s^*+\sum_{s=1}^{n-m}\sigma_{m+s}\rho_s^2h_{s,x}h_s^*\right)$$
$$+{\gamma\over 2}\sum_{s=1}^{n-m}\sigma_{m+s}k_s\rho_s^2(h_sh_s^*-ff^*)+{{\rm i}\hat \lambda\gamma\over 2}ff_x^*.$$
Performing a straightforward computation using (4.10), (4.12), (4.13), (4.15), (4.18) and (4.20)-(4.23) as well as the relations
$$-{{\rm i}\over k_s}\,{\bf z}_{s,x}^*+{\bf z}_s^*=-{\rm i}{\bf z}^*, \quad \sum_{s=1}^{n-m}\sigma_{m+s}\rho_s^2{\bf z}_s={\rm i}\hat\lambda{\bf z}-{\rm i}\tilde{\bf z}_x,$$
$$\sum_{s=1}^{n-m}{\sigma_{m+s}\rho_s^2\over k_s}|D({\bf z}_{s,x}^*; {\bf z})|=\sum_{s=1}^{n-m}\sigma_{m+s}\rho_s^2(|D({\bf z}^*; {\bf z})|-{\rm i}|D({\bf z}_s^*; {\bf z})|),$$
which follow directly from the definition of the vectors ${\bf z}$ and $\tilde{\bf z}$, $\hat S$ can be put into the form
$$\hat S=-\{|D({\bf z}_x^*; {\bf z})|+|D({\bf z}^*; {\bf z}_x)|\}|\bar D|-|D({\bf z}^*; {\bf z})||\bar D({\bf z}^*; {\bf z})|$$
$$-{{\rm i}\gamma\over 2}\Bigl\{|D(\tilde {\bf z}^*; {\bf z}_x)||\bar D|+|\bar D({\bf z}^*; \tilde {\bf z}_x)||D|$$
$$+\sum_{s=1}^{n-m}{\sigma_{m+s}\rho_s^2\over k_s^2}|D({\bf z}_{s}^*; {\bf z}_x)||\bar D({\bf z}^*; {\bf z_s})|+\sum_{s=1}^m\sigma_s|D({\bf a}_s^*; {\bf z}_x)||\bar D({\bf z}^*; {\bf a}_s)|\Bigr\}.$$
In view of the relation (4.25), this expression reduces to
$$\hat S=-(|D({\bf z}_x^*; {\bf z})||\bar D|+|D({\bf z}^*; {\bf z})||\bar D({\bf z}^*; {\bf z})|+|\bar D({\bf z}^*; {\bf z}_x)||D|),$$
which becomes zero by (4.26).\hspace{\fill}$\Box$ \par
\bigskip
\noindent {4.5. One-soliton solutions}\par
\medskip
\noindent  If we put
$$p_1=a+{\rm i}b, \quad z={\rm e}^{a\xi+{\rm i}\{bx+(a^2-b^2-\hat\lambda\gamma b)t\}}, \quad \xi=x-(2b+\hat\lambda\gamma)t, \eqno(4.27a)$$
the tau functions (4.10) for the one-soliton solutions are written in the form
$$f={1\over 2a}\left\{zz^*+{1\over 2}(\mu-\gamma b-{\rm i}\gamma a)c_{11}\right\}, \quad g_j=\alpha_{j1}^*z,\quad j=1, 2, ..., m,$$
$$h_j=f-{zz^*\over a-{\rm i}(b-k_j)}, \quad j=1, 2, ..., n-m. \eqno(4.27b)$$
  The parameter $c_{11}$ from $(4.10d)$ is given  by
$$c_{11}={\sum_{s=1}^m\sigma_s\alpha_{s1}\alpha_{s1}^*\over 
1+{\gamma\over 2}\sum_{s=1}^{n-m}\sigma_{m+s}\rho_s^2\,{-2b+k_s+\mu/\gamma\over a^2+(b-k_s)^2}}. \eqno(4.27c)$$
Note that $c_{11}$ is real. This also follows directly from the Hermitian nature of $c_{jk}$.
If one introduces the real quantities $\beta, \phi$ and $\theta_j$ by (3.30) and puts $(a\beta\gamma)^2={\rm e}^{-4a\xi_0}$, then one can express the
one-soliton solutions compactly in terms of these parameters. After some manipulations, one finds that
$$|q_j|^2={2a^2\alpha_{j1}\alpha_{j1}^*{\rm e}^{2a\xi_0}\over \cosh\,2a(\xi+\xi_0)-{\rm sgn}(a\gamma c_{11})\,\cos\,2\phi},\quad j=1, 2, ..., m, \eqno(4.28)$$
$$|q_{j+m}|^2=\rho_j^2\left[1+{{\rm sgn}(a\gamma c_{11})\{\cos\,2(\theta_j+\phi)+\cos\,2\phi\}\over
   \cosh\,2a(\xi+\xi_0)-{\rm sgn}(a\gamma c_{11})\,\cos\,2\phi}\right],\quad j=1, 2, ...,n- m. \eqno(4.29)$$
The components $q_j$ from (4.28) take the form of bright-solitons with zero background whereas those of (4.29) represent the dark- or bright-solitons with
nonzero background.  
A striking feature of the soliton solutions is that the parameters $a$ and $b$ can be chosen independently unlike the soliton solutions with the pure plane-wave
boundary conditions discussed in section 3.
It turns out that in the case of the mixed type boundary conditions,
the explicit form of the $N$-soliton solution is available without solving algebraic equations.
\par
If one takes the limit $\gamma\rightarrow 0$ under the condition $\mu c_{11}>0$, then the expressions (4.28) and (4.29) reduce to the one-soliton
solutions of the $n$-component NLS system
$$|q_j|^2=a^2\alpha_{j1}\alpha_{j1}^*{\rm e}^{2a\xi_0}\,{\rm sech}^2a(\xi+\xi_0),\quad j=1, 2, ..., m, \eqno(4.30)$$
$$|q_{j+m}|^2=\rho_j^2\left[1 -{a^2\over a^2+(b-k_j)^2}\,{\rm sech}^2a(\xi+\xi_0)\right], \quad j=1, 2, ...,n- m. \eqno(4.31)$$
Note that the condition $\mu c_{11}>0$ assures the regularity of the solutions.  Actually, if $\mu c_{11}<0$, then the solutions exhibit a singularity at $\xi=-\xi_0$. 
 \par
\bigskip
\noindent {\bf Remark 4.2.} The $N$-soliton formula presented in theorem 4.1 have an alternative expressions. Indeed, in accordance with the procedure developed in [13],
we can show that the tau functions given below satisfy the system of bilinear equations (4.4)-(4.7)
$$f=\begin{vmatrix} \hat A & I \\ -I & \hat B \end{vmatrix}, \quad
g_s=\begin{vmatrix} \hat A & I & {\bf z}^T\\ -I & \hat B &{\bf 0}^T \\ {\bf 0} & -{\bf a}_s^* & 0\end{vmatrix}, \ s=1, 2, ..., m,$$
$$h_s=\begin{vmatrix} \hat A & I & {\bf z}^T\\ -I &\hat  B &{\bf 0}^T \\ {\bf z}_s^*/k_s & {\bf 0} & 1\end{vmatrix}, \ s=1, 2, ..., n-m, \eqno(4.32a)$$
$$\hat A=(\hat a_{jk})_{1\leq j,k\leq N},\quad \hat a_{jk}={z_jz_k^*\over p_j+p_k^*}, \quad \hat B=(\hat b_{jk})_{1\leq j,k\leq N},\quad \hat b_{jk}={{1\over 2}(\mu+{\rm i}\gamma p_k)c_{jk}\over p_j+p_k^*}, \eqno(4.32b)$$
where the $N$-component row vectors ${\bf z}, {\bf z}_s$ and ${\bf a}_s$ are defined in (4.10) and $I$ is an $N\times N$ unit matrix.
It is noteworthy that the tau functions $f$ and $g_s$ have the same forms as those of the bright $N$-soliton solution presented in [13]. 
Indeed, if one puts $\rho_s=0\ (s=1, 2, ..., n-m)$, or equivalently $q_{m+s}=0\ (s=1, 2, ..., n-m)$, then $q_s=g_sf^*/f^2\ (s=1, 2, ..., m)$ solve the system of PDEs (1.1)
with the boundary conditions $q_s\rightarrow 0, |x|\rightarrow \infty$.  Since $m$ is an arbitrary positive integer, this gives another proof of the bright $N$-soliton solution. 
\par
\bigskip
\noindent {\bf 5. Concluding remarks}\par
\bigskip
\noindent  The primary advantage of the direct method is that it is capable of  providing a simple mean to obtain soliton solutions irrespective of the boundary conditions.
 The $N$-soliton formulas presented  in this paper  include as special cases the existing
$N$-soliton solutions of the NLS and derivative NLS equations as well  as the Manakov system with nonzero boundary conditions.
The compact expressions of the soliton solutions are particularly useful for investigating their structures, asymptotic behaviors and dynamics. 
On the other hand, the IST is technically involved for nonzero boundary conditions, and hence it
has been applied mainly to the multi-component system such as (1.1) under the restricted class of boundary conditions in which all
the components have the same plane-wave boundary condition at infinity, for instance. 
The development of the IST for the multi-component systems with the general nonzero boundary conditions is still in progress. 
In particular, the derivation of the $N$-soliton solutions by means of the IST is an important issue left to the future.
\par
In conclusion, it will be worthwhile to comment on an integrable system associated with the system (1.1).   The multi-component Fokas-Lenells (FL) system
$$u_{j, xt}=u_j-{\rm i}\left\{\left(\sum_{s=1}^n\sigma_su_{s,x}u_s^*\right)u_j+\left(\sum_{s=1}^n\sigma_su_su_s^*\right)u_{j,x}\right\},\quad u_j=u_j(x,t)\in\mathbb{C},
  \quad j=1, 2, ..., n, \eqno(5.1)$$
is an integrable multi-component generalization  of the FL equation which describes the nonlinear propagation of short pulses in a monomode fiber [36, 37]. 
It belongs to  the first negative flow of the multi-component derivative NLS  hierarchy [38, 39].  The FL equation is a special case of the system (5.1) with $n=1$. 
Its  $N$-soliton solutions have been obtained by employing the direct method for both zero and plane-wave boundary conditions [40, 41].
In view of the above observation on an integrable hierarchy, the structure of the $N$-soliton solution of the multi-component FL system  
is closely related to that of the multi-component derivative NLS system. 
This statement has been confirmed for the one-component system [41].
Thus, the bilinearization and the construction of the $N$-soliton solution of the system (5.1)
with zero and nonzero boundary conditions will be performed in accordance with the procedure developed in the present paper. This interesting issue is currently under study, and the results will be
reported elsewhere. \par
\bigskip
\noindent {\bf Appendix A. Proof of lemma 3.4} \par
\bigskip
\noindent {\bf Proof of (3.27).}  Let $L_1$ be the left-hand side of (3.27). Applying formula (2.3) with ${\bf z}_s$ from $(3.17b)$ and manipulating the resultant expression, one can show that
\begin{align}
L_1 &= \sum_{s=1}^n{\sigma_s\rho_s^2\over k_s^2}|D({\bf z}_s^*; {\bf z}_s)| \notag \\
&=-\sum_{q,r=1}^N{z_qz_r^*\over p_q+p_r^*}D_{qr}\sum_{s=1}^n\sigma_s\rho_s^2\left[{{\rm i}\{{\rm i}(p_q-p_q^*)+k_s\}+p_q\over (p_q-{\rm i}k_s)(p_q^*+{\rm i}k_s)}
-{{\rm i}\{{\rm i}(p_r-p_r^*)+k_s\}-p_r^*\over (p_r-{\rm i}k_s)(p_r^*+{\rm i}k_s)}\right] . \tag{A.1}
\end{align}
In view of (3.16), this expression simplifies to
$$L_1=-\sum_{q,r=1}^ND_{qr}\{\beta_rz_qz_r^*+{\rm i}(\beta_q-\beta_r)d_{qr}\}, \eqno(A.2) $$
where the definition of $d_{jk}$ from $(3.15b)$ and that $\beta_j$ from (3.29) have been used.  Referring to formulas (2.3) and (2.5) ,  $(A.2)$  becomes 
$L_1=|D({\bf z}^*B; {\bf z})|$.   \hspace{\fill}$\Box$\par 
\medskip
\noindent {\bf Proof of (3.28).}  Let $L_2$ be the left-hand side of (3.28).   Use the expansion formula (2.7)  to obtain
\begin{align}
L_2 &= \sum_{s=1}^n{\sigma_s\rho_s^2\over k_s^2}\,|D({\bf z}_s^*, {\bf a}^*; {\bf b}, {\bf z}_s)| \notag \\
&=\sum_{s=1}^n{\sigma_s\rho_s^2\over k_s^2}\sum_{\substack{p,q,j,k=1\\ (p\not=q, j\not=k)}}^Nb_p({\bf z}_s)_q({\bf z}_s^*)_ja_k^*D_{pq,jk}. \tag{A.3}
\end{align}
A similar computation to that leading to  $(A.2)$ gives
\begin{align}
L_2 &= \sum_{\substack{p,q,j,k=1\\ (p\not=q, j\not=k)}}^N b_pa_k^*D_{pq,jk} \{\beta_jz_qz_j^*+{\rm i}(\beta_q-\beta_j)d_{qj}\} \notag  \\
&=\sum_{\substack{p,q,j,k=1\\ (p\not=q, j\not=k)}}^N(\beta_jz_j^*)a_k^*b_pz_qD_{pq,jk} 
-{\rm i}\sum_{p,k=1}^Nb_pa_k^*\left\{\sum_{\substack{q=1\\ (q\not=p)}}^N\beta_qD_{pk}-\sum_{\substack{j=1\\ (j\not=k)}}^N\beta_jD_{pk}\right\} \notag  \\
&=\sum_{\substack{p,q,j,k=1\\ (p\not=q, j\not=k)}}^N(\beta_jz_j^*)a_k^*b_pz_qD_{pq,jk}+{\rm i}\sum_{p,k=1}^Nb_pa_k^*(\beta_pD_{pk}-\beta_kD_{pk}). \tag{A.4}
\end{align} 
Referring to  formulas (2.3) and (2.7) with (2.1), $(A.4)$ can be written in the form
$$L_2=|D({\bf z}^*B, {\bf a}^*; {\bf b}, {\bf z})|
-{\rm i}(|D({\bf a}^*; {\bf b}B)|-|D({\bf a}^*B; {\bf b})|), \eqno(A.5) $$
which is just the right-hand side of (3.28).  \hspace{\fill}$\Box$  \par
 \bigskip
\noindent {\bf Appendix B. Proof of lemma 4.3} \par
\bigskip
 \noindent Let $L_3$ be the left-hand side of (4.24).   In view of formula (2.3) with ${\bf z}_s$ from $(3.17b)$, $L_3$ becomes
 $$L_3=\sum_{s=1}^{n-m}\sigma_{m+s}\rho_s^2\sum_{j,l,q,r=1}^N{z_jz_l^*z_qz_r^*\over p_q+p_l^*}D_{jl}\bar D_{qr}\left({1\over p_l^*+{\rm i}k_s}+{1\over p_q-{\rm i}k_s}\right).$$
 It follows from the definition of $d_{jk}$ from $(4.10b)$ and $\bar d_{jk}$ from (4.20) that
 $${z_qz_l^*\over p_q+p_l^*}=d_{ql}+{{\rm i}\gamma\over 2}\,{p_l^*c_{ql}\over p_q+p_l^*}=\bar d_{ql}-{{\rm i}\gamma\over 2}\,{p_qc_{ql}\over p_q+p_l^*}. $$
 Substituting this expression into $L_3$, using formula (2.5) and referring to formula (2.3), one obtains
 $$L_3=-|D(\tilde {\bf z}^*; {\bf z})||\bar D|-|\bar D({\bf z}^*; \tilde {\bf z})||D|
 -{\gamma\over 2}\sum_{j,l,q,r=1}^N\sum_{s=1}^{n-m}\sigma_{m+s}\rho_s^2\,{{\rm i}(p_q-p_l^*)+k_s\over (p_q-{\rm i}k_s)(p_l^*+{\rm i}k_s)}\,c_{ql}D_{jl}\bar D_{qr}z_jz_r^*. $$
 The relation
 $${\gamma\over 2}\sum_{s=1}^{n-m}\sigma_{m+s}\rho_s^2\,{{\rm i}(p_q-p_l^*)+k_s\over (p_q-{\rm i}k_s)(p_l^*+{\rm i}k_s)}\,c_{ql}
 =-c_{ql}+\sum_{s=1}^m\sigma_s\alpha_{sq}\alpha_{sl}^*, $$
  follows directly from $(4.10d)$, which simplifies the third term of $L_3$ as
 $${\gamma\over 2}\sum_{j,l,q,r=1}^N\sum_{s=1}^{n-m}\sigma_{m+s}\rho_s^2\,{{\rm i}(p_q-p_l^*)+k_s\over (p_q-{\rm i}k_s)(p_l^*+{\rm i}k_s)}\,c_{ql}D_{jl}\bar D_{qr}z_jz_r^*$$
 $$=-\sum_{j,l,q,r=1}^Nc_{ql}D_{jl}\bar D_{qr}z_jz_r^*+\sum_{j,l,q,r=1}^N\sum_{s=1}^m\sigma_s\alpha_{sq}\alpha_{sl}^*D_{jl}\bar D_{qr}z_jz_r^*.$$
 Using the relation $c_{ql}=-(2{\rm i}/\gamma)(\bar d_{ql}-d_{ql})$ from (4.20) and applying formulas (2.3) and (2.5), one can  rewrite  the above expression in terms of bordered determinans. Explicitly, it reads
 $${\gamma\over 2}\sum_{j,l,q,r=1}^N\sum_{s=1}^{n-m}\sigma_{m+s}\rho_s^2\,{{\rm i}(p_q-p_l^*)+k_s\over (p_q-{\rm i}k_s)(p_l^*+{\rm i}k_s)}\,c_{ql}D_{jl}\bar D_{qr}z_jz_r^*$$
 $$=-{2{\rm i}\over\gamma}(|D({\bf z}^*; {\bf z})||\bar D|-|\bar D({\bf z}^*; {\bf z})||D|)
+\sum_{s=1}^{m}\sigma_s|D({\bf a}_s^*; {\bf z})||\bar D({\bf z}^*; {\bf a}_s)|.$$
Last, on substituting this relation into $L_3$, one can see that it coinsides with the right-hand side of (4.24).  The proof of (4.25) parallels completely to that of (4.24). \hspace{\fill}$\Box$ \par
  \bigskip
\noindent {\bf Appendix C. Proof of lemma 4.4} \par
\bigskip
 \noindent Introduce an $N\times N$ matrix $\tilde C=(\tilde c_{jk})_{1\leq j,k\leq N}$ whose elements are defined by the relations 
 $$c_{jk}={2(p_j+p_k^*)z_jz_k^*\over {\rm i}\gamma p_jp_k^*}\left(\tilde c_{jk}-{p_k^*\over p_j+p_k^*}\right), \quad j, k=1, 2, ..., N. $$
 Consequently,  $d_{jk}=(1-\tilde c_{jk})z_jz_k^*/p_j, \bar d_{jk}=\tilde c_{jk}z_jz_k^*/p_k^*$. 
 The determinants associated with the matrices $D$ and $\bar D$ are expressed  by the matrix $\tilde C$ as  follows:
 $$|D|=\kappa|U-\tilde C|, \quad |\bar D|=\kappa|\tilde C|, \quad |D({\bf z}_x^*; {\bf z})|=\kappa \begin{vmatrix} U-\tilde C & {\bf p}^T\\ {\bf p}^* & 0\end{vmatrix},
 \quad |\bar D({\bf z}^*; {\bf z}_x)|=\kappa \begin{vmatrix} \tilde C & {\bf p}^T\\ {\bf p}^* & 0\end{vmatrix},$$
 $$ |D({\bf z}^*; {\bf z})|=\kappa \begin{vmatrix} U-\tilde C & {\bf p}^T\\ {\bf 1} & 0\end{vmatrix},
  \quad |\bar D({\bf z}^*; {\bf z})|=\kappa \begin{vmatrix} \tilde C & {\bf 1}^T\\ {\bf p^*} & 0\end{vmatrix},\quad  \kappa=\prod_{j=1}^N(z_jz_j^*/p_j), \eqno(C.1)$$
 where $U$ is an $N\times N$ matrix whose elements are all unity, and ${\bf p}$ and ${\bf 1}$ are $N$-component row vectors given by
 ${\bf p}=(p_1, p_2, ..., p_N), \ {\bf 1}=(1, 1, ..., 1)$. The above relations are verified by using  a few basic properties of determinants.
 Substituting $(C.1)$ into the left-hand side of (4.26) which is denoted by $L_4$, one finds that
 $$L_4=\kappa\kappa^*\left\{\begin{vmatrix} U-\tilde C & {\bf p}^T\\ {\bf p}^* & 0\end{vmatrix}|\tilde C|
 +\begin{vmatrix} U-\tilde C & {\bf p}^T\\ {\bf 1} & 0\end{vmatrix}\begin{vmatrix} \tilde C & {\bf 1}^T\\ {\bf p}^* & 0\end{vmatrix}
 +\begin{vmatrix} \tilde C & {\bf p}^T\\ {\bf p}^* & 0\end{vmatrix}|U-\tilde C|\right\}. \eqno(C.2)$$
 By an elementary computation, one can show that all the expressions in  $(C.2)$ can be written by the bordered deternminants associated with the matrix $\tilde C$. Explicitly,
 $$\begin{vmatrix} U-\tilde C & {\bf p}^T\\ {\bf p}^* & 0\end{vmatrix}=(-1)^{N-1}(|\tilde C({\bf p}^*; {\bf p})|+|\tilde C({\bf p}^*, {\bf 1}; {\bf p}, {\bf 1})|), 
 \quad \begin{vmatrix} U-\tilde C & {\bf p}^T\\ {\bf 1} & 0\end{vmatrix}=(-1)^{N-1}|\tilde C({\bf 1}; {\bf p})|,$$
 $$|U-\tilde C|=(-1)^N(|\tilde C|+|\tilde C({\bf 1}; {\bf 1})|). \eqno(C.3)$$
 Substitution of (C.3) into (C.2) gives
 $$L_4=(-1)^{N-1}\kappa\kappa^*(|\tilde C({\bf p}^*, {\bf 1}; {\bf p}, {\bf 1})||\tilde C|+|\tilde C({\bf 1}; {\bf p})||\tilde C({\bf p}^*; {\bf 1})|-|\tilde C({\bf p}^*; {\bf p})||\tilde C({\bf 1}; {\bf 1})|).$$
 This expression becomes zero by Jacobi's identity (2.4). \hspace{\fill}$\Box$ \par

\newpage

\leftline{\bf References}\par
\begin{enumerate}[{[1]}]
\item Dodd R K, Eilbeck J C, Gibbon J D and Morris H C 1982 {\it Solitons and Nonlinear Wave Equations} (London: Academic)
\item Novikov S P, Manakov S V, Pitaevskii L P and Zakharov V E  1984 {\it Theory of Solitons:  the Inverse Scattering Method} (New York: Plenum)
\item Ablowitz M J and Clarkson P A 1992 {\it Solitons, Nonlinear Evolution Equations and Inverse Scattering} (Cambridge: Cambridge University Press)
\item Lamb G L {\it Elements of Soliton Theory} 1980 (New York: John Wiley)
\item Rogers C and Shadwick W F 1982 {\it B\"acklund Transformations and Their Applications} (New York: Academic)
\item Matveev V B and Salle M A 1991 {\it Darboux Transformations and Solitons} (Berlin: Springer)
\item Hirota R 1980 { Direct methods in soliton theory} {\it Solitons (Topics in Current Physics)} ed R K Bullough and P J Caudrey (Berlin: Springer) pp 157-75
\item Matsuno Y 1984 {Bilinear Transformation Method} (New York:  Academic)
\item Hirota R 2004 {The direct Method in Soliton Theory} (Cambridge: Cambridge University Press)
\item Hasegawa A and Kodama Y 1995 {\it Solitons in Optical Communications } (New York: Oxford)
\item Kivshar Y S and Agrawal G P 2003 {\it Optical Solitons From Fibers to Photonic Crystals} (New York: Academic)
\item Maimistov A I 2010 { Solitons in nonlinear optics} {\it Quantum Electron.} {\bf 40} 756-81
\item Matsuno Y 2011 {The bright $N$-soliton solution of a multi-component modified nonlinear Schr\"odinger equation} {\it J. Phys. A: Math. Theor.} {\bf 44} 495202
\item Hisakado M and Wadati M 1995  Integrable multi-component hybrid nonlinear Schr\"odinger equations {\it J. Phys. Soc. Japan} {\bf 64} 408-13
\item Zakharov V E and Shabat A B 1973 Interaction between solitons in a stable medium {\it Sov. Phys.-JETP} {\bf 37} 823-8
\item Faddeev L D and Takhtajan L A 1987 {\it Hamiltonian Methods in the Theory of Solitons} (Berlin, Heidelberg and New York: Springer)
\item Gelash A A and Zakharov V E 2014 {Superregular solitonic solutions; a novel scenario for the nonlinear stage of modulation instability} {\it Nonlinearity} {\bf 27} R1-R39
\item Biondini G and Kova\v{c}i\v{c} G 2014 Inverse scattering transform for the focusing nonlinear Schr\"odinger equation with nonzero boundary conditions {\it J. Math. Phys.} {\bf 55} 031506
\item Kawata T and Inoue H 1978 {Exact solutions of the derivative nonlinear Schr\"odinger equation under the nonvanishing conditions} {\it J. Phys. Soc. Jpn.} {\bf 44} 1968-76
\item Ieda J, Uchiyama M and Wadati M 2007 {Inverse scattering method for square matrix nonlinear Schr\"odinger equation under nonvanishing boundary conditions} {\it J. Math. Phys.} {\bf 48} 013507
\item Prinari B, Biondini G and Trubatch A D  2011 {Inverse scattering transform for the multi-component nonlinear Schr\"odinger equation with nonzero boundary conditions} {\it Stud. Appl. Math.} {\bf 126} 245-302
\item  Dubrovin B A, Malanyuk T M, Krichever I M and Makhan'kov V G 1988 { Exact solutions of the time-dependent Schr\"odinger equation
           with self-consistent potentials} {\it Sov. J. Part. Nucl.} {\bf 19} 252-269
\item Park Q.-H and Shin H J 2000 Systematic construction of multicomponent optical solitons {\it Phys. Rev. E} {\bf 61} 3093-106
\item Park Q.-H and Shin H J 2001 Darboux transformation and Crum's formula for multi-component integrable equations {\it Physica D} {\bf 157} 1-15
\item Ling L, Zhao L.-C and Guo B 2015 Darboux transformation and multi-dark soliton for $N$-component coupled nonlinear Schr\"odinger equations {\it Nonlinearity} {\bf 28} 3243-61
\item Gadzhimuradov T A, Abdullaev G O and Agalarov A M 2017 Vector dark solitons with oscillating background density {\it Nonlinear Dyn.} {\bf 89} 2695-702
\item Tsuchida T 2013 {Exact solutions of multicomponent nonlinear Schr\"odinger equations under general plane-wave boundary conditions} {\it arXiv: 1308.6623v2}
\item Manakov S V 1974 { On the theory of two-dimensional stationary self-focusing of electromagnetic waves} {\it Sov. Phys. - JETP} {\bf 38} 248-253
\item  Vijayajayanthi M, Kanna T and Lakshmanan M  2009 {Multisoliton solutions and energy sharing collisions in coupled nonlinear Schr\"odinger equations with
       focusing, defocusing and mixed type nonlinearities} {\it Eur. Phys. J. Special Topics} {\bf 173} 57-80
\item Feng B.-F 2014 General $N$-soliton solution to a  vector nonlinear Schr\"odinger equation {\it J. Phys. A: Math. Theor.} {\bf 47} 355203
\item Ohta Y, Wang D.-S and Yang J 2011 General $N$-dark-dark solitons in coupled nonlinear Schr\"odinger equations {\it Stud. Appl. Math.} {\bf 127} 345-71
\item Zhang H.-Q, Tian B,  L\"u X, Li He and Meng X.-H 2009 Soliton interaction in the coupled mixed derivative nonlinear Schr\"odinger equations {\it Phys. Lett. A} {\bf 373} 4315-21
\item Matsuno Y 2011 The $N$-soliton solution of a two-component modified nonlinear Schr\"odinger equation {\it Phys. Lett. A} {\bf 375} 3090-94
\item Vein R and Dale P 1999 {\it Determinants and Their Applications in Mathematical Physics} (New York: Springer)
\item Tajiri M and Watanabe Y 1998 Breather solutions to the focusing nonlinear Schr\"odinger equation {\it Phys. Rev. E} {\bf 57} 3510-19
\item  Fokas A S 1995 On a class of physically important integrable equations {\it Physica } D {\bf 87} 145-50
\item  Lenells J 2009 Exactly solvable model for nonlinear pulse propagation in optical fibers {\it Stud. Appl. Math.} {\bf 123} 215-32 
\item  Fordy A P 1984  Derivative nonlinear Schr\"odinger equations and Hermitian symmetric spaces {\it J. Phys. A: Math. Gen.} {\bf 17}  1235-45
\item Tsuchida T and Wadati M 1999  New integrable systems of derivative nonlinear Schr\"odinger equations with multiple components {\it Phys. Lett.} A {\bf 257} 53-64
\item  Matsuno Y  2011 A direct method of solution for the Fokas-Lenells derivative nonlinear Schr\"odinger equation: I. Bright soliton solutions {\it J. Phys. A: Math. Theor.} {\bf 44}  235202
\item  Matsuno Y  2012 A direct method of solution for the Fokas-Lenells derivative nonlinear Schr\"odinger equation: II. Dark soliton solutions {\it J. Phys. A: Math. Theor.} {\bf 45}  475202

\end{enumerate}

\end{document}